\documentclass{amsart}

\usepackage[american]{babel}
\usepackage{csquotes}

\usepackage{csquotes,amsmath, amsthm, amssymb, amsbsy, mdwlist}

\usepackage[belowskip=-5pt,aboveskip=5pt]{caption}
\usepackage[hyperfigures=true]{hyperref}

\usepackage{ifthen}
\usepackage[usenames]{color}
\newcounter{content}
\newcounter{notes}
\newcounter{techreport}

\usepackage{amssymb,amsmath,amsthm}
\usepackage{graphicx}
\usepackage{mathrsfs} 
\usepackage{amsmath}
\usepackage{amssymb}
\usepackage[all]{xy}
\usepackage{color, graphics}
\usepackage{palatino, url, multicol}
\usepackage{subfigure}
\usepackage{mathrsfs}
\usepackage{multicol}
\usepackage{tabularx}

\newcommand{\bi}{\begin{itemize}}
\newcommand{\ei}{\end{itemize}}
\definecolor{Emphcolor}{cmyk}{0,0.89,0.94,0.1}
\definecolor{Netcolor}{rgb}{.8,0,.9}
\definecolor{Diseasecolor}{rgb}{1,.8,.2}
\definecolor{Sampcolor}{rgb}{0,.9,.3}
\definecolor{Black}{rgb}{0,0,0}
\definecolor{Red}{rgb}{1,0,0}
\definecolor{Blue}{rgb}{0,0,1}
\definecolor{Gray}{gray}{.6}

\font\elevenrm=cmr11

\newcommand{\qrns}{{\elevenrm "}\kern-2pt}

\theoremstyle{plain}
\newtheorem{thm}{Theorem}[section]
\newtheorem{cor}[thm]{Corollary}
\newtheorem{lem}[thm]{Lemma}

\theoremstyle{definition}

\usepackage{float}
\usepackage{rotating}
\usepackage{natbib}
\usepackage{color}
\usepackage{hypernat}
\usepackage{pdflscape}
\usepackage{fullpage}

\usepackage{graphics}
\usepackage{subfigure}
\usepackage[all]{xy}
\usepackage{amsmath}
\usepackage{amssymb}
\usepackage[all]{xy}
\usepackage{palatino, url, multicol}
\usepackage{mathrsfs}
\usepackage{multicol}
\usepackage{enumerate}

\newcommand{\mc}[1]{\mathcal{#1}}
\newcommand{\mbb}[1]{\mathbb{#1}}
\newcommand{\h}[1]{\hat{#1}}
\newcommand{\fd}{f_1\left(v,d,\theta_0,0\right)}
\newcommand{\fp}{f_2\left(v,\bp,\theta_0,0\right)}

\newcommand{\dfd}{\partial f_1\left(v,d,\theta,\lambda\right)/\partial\left(\theta,\lambda\right)}
\newcommand{\dfp}{\partial f_2\left(v,\bp,\theta,\kappa\right)/\partial\left(\theta,\kappa\right)}
\newcommand{\dfdt}{\partial f_1\left(v,d,\theta_0,0\right)/\partial\left(\theta,\lambda\right)}
\newcommand{\dfpt}{\partial f_2\left(v,\bp,\theta_0,0\right)/\partial\left(\theta,\kappa\right)}
\newcommand{\norm}[1]{\left|\left|#1\right|\right|}
\newcommand{\mG}{\mathbb{G}}
\newcommand{\mK}{\mathbb{K}}
\newcommand{\mH}{\mathbb{H}}

\newcommand{\ind}{\perp\!\!\!\perp}
\newcommand{\cind}[3]{#1\ind#2\mid#3}
\newcommand{\bp}{\nu}
\newcommand{\Bp}{{\scriptstyle\Upsilon}}
\newcommand{\pf}{F_{\mc{S}}}
\newcommand{\argmax}{\arg~\max}

\title[]{Population level information combined parameter estimation from complex survey datasets}
\author{Sanjay Chaudhuri}
\address{Department of Statistics and Applied Probability, National University of Singapore, Singapore $117546$}
\email{stasc@nus.edu.sg}
\author{Mark S. Handcock}
\address{Department of Statistics, University of California, Los Angeles}
\email{handcock@ucla.edu}
\author{Michael S. Rendall}
\address{Department of Sociology, University of Maryland College Park}
\email{mrendall@umd.edu}

\date{\today} 

\begin{document}

\begin{abstract}
We consider an empirical likelihood framework for
inference for a statistical model 
based on an informative sampling design and population-level information.
The population-level information is summarized in the form of the estimating
equations and incorporated in the inference through additional constraints.
Covariate information is incorporated both through the weights and 
the estimating equations.
The estimator is based on conditional weights.
We show that under usual conditions, with population size increasing unbounded, the estimates are strongly consistent,
asymptotically unbiased and normally distributed.  Moreover, they
are more efficient than other probability weighted analogues.  Our framework provides
additional justification for inverse probability weighted score
estimators in terms of conditional empirical likelihood.  We give
an application to demographic hazard modeling by combining birth registration data with panel survey data to estimate annual first birth probabilities.

\end{abstract}

\maketitle
\section{Introduction}

In many applications in statistics and the social sciences, the use of population level information and sample survey data in conjunction is beneficial.  In sample surveys, data are collected for large number of variables, and thus meaningful models for the behavior of a response of interest can be specified.  
  However, survey data suffer from sampling error and from bias due to non-response. On the other hand, population level data collected from, e.g., census and vital events registration systems, typically do not contain a sufficient range of variables
 to specify meaningful models, but are collected with comparatively less error and are less biased.  
These complementary strengths and weaknesses suggest that a combination of population and sample data may produce more meaningful and efficient estimates of the model parameters and lead to better inference.


Several methods for incorporating population level information in sample-based modeling have been investigated.  
One procedure is to express the population level information as functions of the model parameters and to use them as restrictions in parameter estimation. 
 \citet{mshren1,mshren2} consider a \emph{constrained maximum likelihood estimator} (CMLE) while \citet{imbens2} use \emph{generalized method-of-moments} (GMM) to incorporate the constraints.  It is known that both these methods produce asymptotically normal and unbiased estimates.  Analytic forms of their asymptotic covariance matrices are known. 
However, both methods, particularly the CMLE, are computationally intensive and generally difficult to handle.  The likelihood is constrained by non-linear equality constraints, and  
though standard software to perform such optimization exist, optimization may be prohibitively slow for even moderate numbers of explanatory variables and population level constraints.

Empirical likelihood, introduced by \citet{owenbook}, provides a semi-parametric method for augmentation of the population level information.  Prior to this, \citet{hartley_rao_1968,hartley_rao_1969} used similar techniques in survey sampling under the name of \emph{scale-load} approach. 
\citet{Qinlawless1} showed that the empirical likelihood can be used to define a \emph{profile empirical likelihood} of the parameters, which can then be maximized to obtain the parameter estimates.  They further showed that these estimates are asymptotically unbiased and normally distributed. 
\citet{chen_qin_1993} and \citet{chen_sitter_1999} used an empirical likelihood based method to incorporate auxiliary information available in a sample drawn from a finite population. 
  For equal probability sampling, \citet{chaudhuri_handcock_rendall_2008} 
used an empirical likelihood based method to augment population level information in sample-based generalized linear modeling and developed a simple two-step method to estimate the model parameters. 
 Under standard regulatory conditions, this two-step estimator is strongly consistent, asymptotically normal and unbiased. Further, incorporation of population level information reduces the standard error of the parameter estimates.  
Empirical likelihood based methods maximize their objective function under linear equality constraints and so are computationally much less demanding than the CMLE. Unlike the CMLE, no parametric form of the distribution needs to be specified and so it is much more flexible and easy to implement. 
Moreover, the efficiency of this estimator is very close to the CMLE under the correctly specified parametric model and usually much better than the CMLE under misspecified models \citep{chaudhuri_drton_richardson_biom_2007}.          


Real-world surveys are complex, with observations drawn according to
informative designs and are accompanied by weights.
In this paper we investigate methods of combining the
inclusion-probability contained in these design weights with sampled
observations and population level information through empirical
likelihood based methods.


Empirical likelihood based methods which take into account the
design weights in the sample have been studied by several
researchers.  \citet{chen_sitter_1999}, motivated by the
Horvitz-Thompson estimator in survey sampling, proposed a
\emph{pseudo-empirical likelihood}.  Wu and colleagues (notably
\citet{chen_sitter_wu_2002}, \citet{wu_rao_2006},
\citet{rao_wu_2008} among others) study this method extensively
and apply it to several design based surveys.  The pseudo
empirical likelihood can be re-interpreted as a ``backward''
Kullback-Leibler divergence of the unknown weights from the
sampling weights.  The distribution is specified by the choosing
the weights that minimize this divergence.  \citet{wu_2004}
discuss a similar minimized weighted entropy estimator.

In this paper we develop a framework that produces a 
different procedure to the above. We use the framework to compare the two procedures (Section 6). 
We consider the conditional distribution of the sample given that they were selected in the sample and estimate their distribution in the population.  \citet{rao_patil_1978} considered a similar but more restrictive parametric approach and implemented it on size-biased sampling.  
Pfeffermann and colleagues (e.g. \citet{pfeffermann_1998}, \citet{pfeffermann_sverchkov_1999}, \citet{krieger_pfeffermann_1992}) investigated its use in parametric modeling of survey data.
  Non-parametric estimators of the population distribution using the same principles were investigated by \citet{vardi_1985}.  He considered multiple samples drawn from a population through different designs and provided conditions for the existence and uniqueness of the non-parametric estimator of the population distribution. 
The asymptotic properties of this non-parametric estimator have been studied by \citet{gill_vardi_wellner_1988}.  Non-parametric estimation under the same model and design has been studied by \citet{chambers_dorfman_sverchkov_2003}.  \citet{qin_1993} employed empirical likelihood in a two-sample testing problem, where only one sample is biased by the design.  
  He showed that under certain conditions the empirical log-likelihood ratio has an asymptotic Chi-squared limit.  A similar approach has been taken by \citet{qin_leung_shao_2002} to analyze data with non-ignorable non-response. 
  \citet{qin_zhang_2007} use an empirical likelihood based method in observational studies where part of the response is missing.  Calibration estimation using a similar empirical likelihood in Poisson sampling has been considered by \citet{kim_2009}.


We develop an empirical likelihood method based on the conditional likelihood used by Pfeffermann and colleagues (e.g., \citet{pfeffermann_1998}).  Population and model information are incorporated to infer from a sample drawn according to a stratified design. 
In Section \ref{sec:gen} we interpret the sampling weights as random variables depending on all observations of the design variables in the population. 
We also introduce our model and develop a composite likelihood to incorporate sampling weights in our analysis.
In Section \ref{sec:emp} we use empirical likelihood and incorporate population level information in our composite likelihood.
  A two-step estimation procedure to estimate the model parameters by maximizing the empirical likelihood is developed in  Section \ref{sec:comp}.
The asymptotic properties of this composite empirical likelihood based estimator are compared with the pseudo-empirical likelihood estimator of \citet{chen_sitter_1999}  (CS) and the unconstrained parametric \emph{pseudo likelihood} (PL) estimator.  
We show that, under standard regularity conditions, the composite empirical likelihood (CE) estimator is strongly consistent, asymptotically unbiased and has an asymptotic normal limit (Section \ref{sec:asymp}). 
The CE estimator is shown to be more efficient that the CS and PL estimators in an application to demographic hazard modeling with a complex longitudinal survey (Section \ref{sec:data}). 

\section{Model and Design Specification}\label{sec:gen}

We consider a ``superpopulation'' model with response $Y$, a set of auxiliary variables $X=\left\{X^{(1)},X^{(2)},\ldots,X^{(p)}\right\}$ and a set of design variables $D=\left\{D^{(1)},D^{(2)},\ldots,D^{(q)}\right\}$.
The population is comprised of $N$ i.i.d. draws from the super-population model. We label the elements of the population by $\mc{P}=\{1,2,\ldots,N\}.$ 

A random sample $\mc{S}$ of $n$ observations is drawn from $\mc{P}$ according to a design depending on $D$ and possibly on some unknown parameters (specified in Section \ref{sec:wtdemp}). The available data does not contain all variables in $D$, only a subset $Z=\{Z^{(1)},Z^{(2)},\ldots,Z^{(m)}\}$ is supplied. Let $Z^c=D\setminus Z$.  
Variables in $X$ and $Y$ are not directly involved in the sampling design.  We denote $V=Y\cup X\cup Z$ to be the $m+p+1$ dimensional random vector observed in the dataset.  Further, we collect all the explanatory variables in the model in a set $A\subseteq V$.  
Suppose, $D_{\mc{P}}$, $X_{\mc{P}}$, $Y_{\mc{P}}$, $Z_{\mc{P}}$, $Z^c_{\mc{P}}$ denote the vectors and matrices of all observations of the corresponding variables on the population $\mc{P}$.
For $S\subseteq\mc{P}$, $V_{S}$ is the matrix of observations in $S$. $V_{\bar{S}}$ are the observations not in $S$, where $\bar{S} = \mc{P} \setminus S$.  


Primary scientific intertest focuses on the relationship between a response $Y$
and the set of explanatory variables $A$.
Examples of such models are generalized linear models (GLM)
\citep{mccneldrbook}.
As an important special case, we consider joint models for $Y$ and $A$, $P_{\theta}(Y,A),$ parametrized by $\theta$.
For example, for GLM $\mu(\theta) = A\theta$. We specify the broader class of
applicable models in Section 2.1.
\subsection{Model specification} Suppose $F^0$ is the distribution of $V_1$ in the population with density $dF^0$ w.r.t. a suitable measure. The relationship between the response $Y$ and the set of auxiliary variables $A$ is assumed to be specified by:
\begin{equation}\label{eq:parconstr}
E_{F^0}\left[\psi_{\theta}\left(Y_1,A_1\right)\right]=0.
\end{equation}   
where $\psi$ is a known function depending only on $Y_1$ and $A_1$ and some unknown parameter $\theta.$
There may be several choices for $\psi$ \citep{Qinlawless1}. For parametric models, such as the GLM
considered in the introduction, 
the corresponding \emph{score functions} $S_{\theta}\left(Y,A\right)$ are
natural choices.

Further, certain parameters in the superpopulation may be known without any error. Suppose $g$ is a given functional of $V$ not depending on $\theta$ and
\begin{equation}\label{eq:popconstr}
E_{F^0}\left[g\left(V_1\right)\right]=\gamma.
\end{equation}    
We then say that \eqref{eq:popconstr} specifies \emph{population-level auxiliary information} if $\gamma$ is known without any error.
  
\subsection{Design Specification} 


The sample $\mc{S}$ is a random subset or random multiset for sampling with replacement, of size $n$ of $\mc{P}$.  Below we consider only the subsets of $\mc{P}$, the description for multisets is similar.

For $S\subseteq\mc{P}$ suppose $I_{S}$ is the random indicator function for $S\subseteq\mc{S}$. We assume that $I_S$ satisfy the following conditions,
\begin{enumerate}
\item $I_{\emptyset}=0$, where $\emptyset$ denotes the empty set.
\item For any $S\subseteq\mc{P}$, if $I_S=1$, then $I_T=1$, for all $T\subseteq S$, $T\ne\emptyset$. 
\item For any $S_1,S_2\subseteq\mc{P}$, if $I_{S_1}=1$ and $I_{S_2}=1$, then $I_{S_1\cup S_2}=1$. 
\end{enumerate}
The sample $\mc{S}$ is the unique largest subset $S$ of $\mc{P}$ such that $I_{S}=1$.  Notice that the above definition is valid for any sampling scheme producing samples of fixed or random sample sizes.  
Furthermore, the sample $\mc{S}$ can equivalently be specified by the
set $\{I_S:S\subseteq \mc{P}\}$.

The sampled units are drawn according to a design depending on $D_{\mc{P}}$.  For any $S\subseteq\mc{P}$, the design specifies the conditional probability of $I_S=1$, given $D_{\mc{P}}$.  
Suppose $\pi_S=Pr_{\mc{P}}\left(I_S=1\mid D_{\mc{P}}\right)$, where $Pr_{\mc{P}}\left(\cdot\right)$ is the probability under the population. Thus the $\pi_S$ are represent the joint inclusion probabilities. Notice that, $\pi_S$ is a random variable because of $D_{\mc{P}}$.

If the sample units are drawn according to a design, the sampling mechanism may not be \emph{ignorable}. 
The observed distribution of $V$ in the sample $\mc{S}$ may be different from its distribution in the population and may depend on the particular sample selected. 


\noindent{\bf Assumption $1$: Conditional independence given the design}. For all possible $S\subseteq \mc{P}$, under the population distribution,
$\pi_{S}$ is conditionally independent of $Y_{\mc{P}}$ and $X_{\mc{P}}$ given $D_{\mc{P}}$. That is
\begin{equation}
\cind{\pi_{\mc{S}}}{\left(Y_{\mc{P}},X_{\mc{P}}\right)}{D_{\mc{P}}}.
~~~~~~~~\mbox{for~all}~ S\subseteq\mc{P}.
\end{equation}  

This says that $\pi_S$ depends on the $Y_{\mc{P}}$ and $X_{\mc{P}}$ only through the design variables.
Under Assumption $1$ for all $S$, $E_{\mc{P}}\left[\pi_S\mid Y_{\mc{P}},X_{\mc{P}}, D_{\mc{P}}\right]=E_{\mc{P}}\left[\pi_S\mid D_{\mc{P}}\right]=\pi\left(S,D_{\mc{P}}\right)$, for some function $\pi$.  Now using, $\pi_S=E_{\mc{P}}\left[I_S\mid D_{\mc{P}}\right]$ we obtain the following result. 
\begin{lem}\label{lem:a1}  
Assumption $1$ holds iff $\pi_S=\pi\left(S,D_{\mc{P}}\right)$.
\end{lem}


Lemma \ref{lem:a1} follows from the definition of conditional independence \citep{lau}.  It further shows that, under Assumption $1$, conditioning on $D_{\mc{P}}$ and the pair $\left(D_{\mc{P}},\pi_S\right)$ is same.  The following relationships can also be obtained from Assumption $1$ and Lemma \ref{lem:a1}.  
 
\begin{lem}\label{lem:a2} 
Under Assumption $1$, for all $S\subseteq\mc{P}$, the following holds
\begin{enumerate}
\item $E_{\mc{P}}\left[I_S\mid \pi_S\right]=\pi_S$ {\rm~~~~ and}
\item $\cind{I_S}{D_{\mc{P}}}{\pi_S}$.
\end{enumerate}
\end{lem}
\citet{pfeffermann_1998} use the relationship in Conclusion $1.$ to justify
their parametric likelihood.  The conditional independence relation in $2.$ is
exactly the ``Condition $1$'' in \citet{sugden_smith_1984}, which implies that
under Assumption $1$ the selection procedure (i.e., the actual dependence of
$\pi_{S}$ on $D_{\mc{P}}$) can be ignored for inference given the design. 

\noindent{\bf Assumption $2$: Conditional independence given the sampling probabilities.} We assume that for all $S\subseteq\mc{P}$, under the population distribution, $I_S$ is conditionally independent of $X_{\mc{P}}$, $Y_{\mc{P}}$ and  $D_{\mc{P}}$ given $\pi_{S}$. That is,
\begin{equation}
\cind{I_S}{\left(X_{\mc{P}},Y_{\mc{P}},D_{\mc{P}}\right)}{\pi_{S}}
~~~~~~~~\mbox{for~all}~ S\subseteq\mc{P}.
\end{equation}

Assumption $2$ does not follow from Assumption $1$ except in some special cases.
It says that inclusion depends on the $X_{\mc{P}},Y_{\mc{P}}$ and $D_{\mc{P}}$ only only through the joint inclusion probabilities.
In particular, it implies that the conditional probability of $I_S=1$ given
$X_{\mc{P}}$, $Y_{\mc{P}}$, $D_{\mc{P}}$ and $\pi_{S}$ equals
$E_{\mc{P}}\left[I_S\mid \pi_S\right]=\pi_S$.  \citet{pfeffermann_1998} make
this assumption without stating it explicitly. 

\begin{lem}\label{lem:conda2} For all $S\subseteq\mc{P}$,  under Assumptions $1$, Assumption $2$ is equivalent to $\cind{I_S}{D_{\mc{P}}}{\pi_S}$ and $\cind{I_S}{\left(X_{\mc{P}},Y_{\mc{P}}\right)}{D_{\mc{P}}}$.
\end{lem}
The condition $\cind{I_S}{\left(X_{\mc{P}},Y_{\mc{P}}\right)}{D_{\mc{P}}}$ is the basic design assumption of \citet{scott_1977}.  According to \citet{sugden_smith_1984}, any design which only depends on $D_{\mc{P}}$ should satisfy this condition. 
\begin{lem}\label{lem:conda3} For all $S\subseteq\mc{P}$, Assumptions $1$ and $2$ imply the following conditional independence relationships.
\begin{enumerate}
\item $\cind{I_S}{V_S}{\pi_S}$,
\item $\cind{\left(I_S,\pi_S\right)}{\left(X_{\mc{P}},Y_{\mc{P}}\right)}{D_{\mc{P}}}$ and
\item $\cind{I_S}{\left(X_S,Y_S\right)}{D_{\mc{P}}}$.
\end{enumerate}
\end{lem}

The statement $1.$ of Lemma \ref{lem:conda3} implies that $Pr_{\mc{P}}\left[I_S\mid V_S,\pi_S\right]=Pr_{\mc{P}}\left[I_S\mid\pi_S\right]=\pi_S$ for all $S\subseteq\mc{P}$. From this, following \citet{pfeffermann_sverchkov_2003} we obtain
\begin{align}
Pr_{\mc{P}}\left[I_S=1\mid V_S\right]&=E_{\mc{P}}\left[I_S\mid V_S\right]=E_{\mc{P}}\left[E_{\mc{P}}\left[I_S\mid V_S,\pi_S\right]\mid V_S \right]=E_{\mc{P}}\left[\pi_S\mid V_S \right],\nonumber\\
Pr_{\mc{P}}\left[I_S=1,V_{S}\right]&=Pr_{\mc{P}}\left[I_S=1\mid V_{S}\right]Pr_{\mc{P}}\left[V_{S}\right]=E_{\mc{P}}\left[\pi_S\mid V_S \right]Pr_{\mc{P}}\left[V_{S}\right].\label{eq:joint1}
\end{align}
Note that, Assumption $2$ is sufficient but not necessary for \eqref{eq:joint1} to hold. One of the conditions $\cind{I_S}{V_S}{\pi_S}$ or $\cind{I_S}{\left(X_S,Y_S\right)}{D_{\mc{P}}}$ would suffice.   We could have alternatively assumed:

\noindent{\bf Assumption $2^{\prime}$: } We assume that for all $S\subseteq\mc{P}$, under the population distribution, $I_S$ is conditionally independent of $X_{S}$ and $Y_S$ given $D_{\mc{P}}$.  That is,
\begin{equation}
\cind{I_S}{\left(X_{S},Y_S\right)}{D_{\mc{P}}}
~~~~~~~~\mbox{for~all}~ S\subseteq\mc{P}.
\end{equation}

Unlike Assumption $2$, Assumption $2^{\prime}$ still allows $I_S$ to be conditionally dependent on $\left(X_{\bar{S}},Y_{\bar{S}}\right)$ given $\pi_S$ without violating Lemma \ref{lem:a2}.  This will happen in very special situations where
typically the information about the design available from $\pi_S$ is incomplete and the design is potentially mis-specified. 
We will use Assumption $2$ to exclude these situations.

\begin{figure}[t]
\begin{center}
\setlength{\unitlength}{3947sp}%
\begingroup\makeatletter\ifx\SetFigFont\undefined%
\gdef\SetFigFont#1#2#3#4#5{%
  \reset@font\fontsize{#1}{#2pt}%
  \fontfamily{#3}\fontseries{#4}\fontshape{#5}%
  \selectfont}%
\fi\endgroup%
\begin{picture}(5424,2247)(2089,-2173)
\put(6001,-61){\makebox(0,0)[lb]{\smash{{\SetFigFont{12}{14.4}{\rmdefault}{\mddefault}{\updefault}{$Z^c_{\bar{\mc{S}}}$}%
}}}}
\thinlines
{\put(3601,-136){\vector(-2,-3){1015.385}}
}%
{\put(2476,-1036){\vector( 0,-1){600}}
}%
{\put(3451,-886){\vector(-1, 0){900}}
}%
{\put(3526,-961){\vector(-4,-3){1020}}
}%
{\put(3751,-136){\vector( 1,-2){450}}
}%
{\put(3751,-886){\vector( 4,-3){384}}
}%
{\put(5926,-61){\vector(-4,-3){1548}}
}%
{\put(5926,-886){\vector(-4,-1){1570.588}}
}%
{ \put(3751,-886){\vector( 4,-1){1552.941}}
}%
{ \put(4276,-1336){\vector( 0,-1){300}}
}%
{ \put(6076,-136){\vector( 0,-1){600}}
}%
{ \put(3676,-136){\vector( 0,-1){600}}
}%
{ \put(6151,-61){\vector( 4,-3){984}}
}%
{ \put(6226,-886){\vector( 1, 0){975}}
}%
{ \put(6151,-136){\vector( 2,-3){1015.385}}
}%
{ \put(7276,-1036){\vector( 0,-1){600}}
}%
{ \put(6151,-961){\vector( 4,-3){984}}
}%
{ \put(6001,-136){\vector(-1,-2){495}}
}%
{ \put(5926,-961){\vector(-3,-2){363.462}}
}%
{ \put(5476,-1261){\vector( 0,-1){450}}
}%
{ \put(3751, 14){\vector( 4,-3){1548}}
}%
{ \put(2101,-2161){\dashbox{60}(1800,1500){}}
}%
{ \put(5776,-2161){\dashbox{60}(1725,1500){}}
}%
\put(2401,-961){\makebox(0,0)[lb]{\smash{{\SetFigFont{12}{14.4}{\rmdefault}{\mddefault}{\updefault}{$X_{\mc{S}}$}%
}}}}
\put(3601,-961){\makebox(0,0)[lb]{\smash{{\SetFigFont{12}{14.4}{\rmdefault}{\mddefault}{\updefault}{$Z_{\mc{S}}$}%
}}}}
\put(2401,-1861){\makebox(0,0)[lb]{\smash{{\SetFigFont{12}{14.4}{\rmdefault}{\mddefault}{\updefault}{$Y_{\mc{S}}$}%
}}}}
\put(3601,-1861){\makebox(0,0)[lb]{\smash{{\SetFigFont{12}{14.4}{\rmdefault}{\mddefault}{\updefault}{$V_{\mc{S}}$}%
}}}}
\put(4201,-1261){\makebox(0,0)[lb]{\smash{{\SetFigFont{12}{14.4}{\rmdefault}{\mddefault}{\updefault}{$\pi_{\mc{S}}$}%
}}}}
\put(4201,-1861){\makebox(0,0)[lb]{\smash{{\SetFigFont{12}{14.4}{\rmdefault}{\mddefault}{\updefault}{$I_{\mc{S}}$}%
}}}}
\put(5401,-1261){\makebox(0,0)[lb]{\smash{{\SetFigFont{12}{14.4}{\rmdefault}{\mddefault}{\updefault}{$\pi_{\bar{\mc{S}}}$}%
}}}}
\put(5401,-1861){\makebox(0,0)[lb]{\smash{{\SetFigFont{12}{14.4}{\rmdefault}{\mddefault}{\updefault}{$I_{\bar{\mc{S}}}$}%
}}}}
\put(6001,-961){\makebox(0,0)[lb]{\smash{{\SetFigFont{12}{14.4}{\rmdefault}{\mddefault}{\updefault}{$Z_{\bar{\mc{S}}}$}%
}}}}
\put(7201,-961){\makebox(0,0)[lb]{\smash{{\SetFigFont{12}{14.4}{\rmdefault}{\mddefault}{\updefault}{$X_{\bar{\mc{S}}}$}%
}}}}
\put(7201,-1861){\makebox(0,0)[lb]{\smash{{\SetFigFont{12}{14.4}{\rmdefault}{\mddefault}{\updefault}{$Y_{\bar{\mc{S}}}$}%
}}}}
\put(6001,-1861){\makebox(0,0)[lb]{\smash{{\SetFigFont{12}{14.4}{\rmdefault}{\mddefault}{\updefault}{$V_{\bar{\mc{S}}}$}%
}}}}
\put(3601,-61){\makebox(0,0)[lb]{\smash{{\SetFigFont{12}{14.4}{\rmdefault}{\mddefault}{\updefault}{$Z^c_{\bar{\mc{S}}}$}%
}}}}
{ \put(3526,-61){\vector(-4,-3){984}}
}%
\end{picture}%
\vskip 12pt
\caption{A graphical representation of Assumption $1$ through \emph{directed acyclic graphs}.  It represents that $\cind{I_{\mc{S}}}{\left(Z_{\mc{P}},Z^c_{\mc{P}}\right)}{\pi_{\mc{S}}}$ and $\cind{\left(I_{\mc{S}},\pi_{\mc{S}}\right)}{\left(Y_{\mc{P}},X_{\mc{P}}\right)}{D_{\mc{P}}}$ holds.  The directed edges do not necessarily indicate a causal relationship.}
\label{fig:graph}
\end{center}
\end{figure}
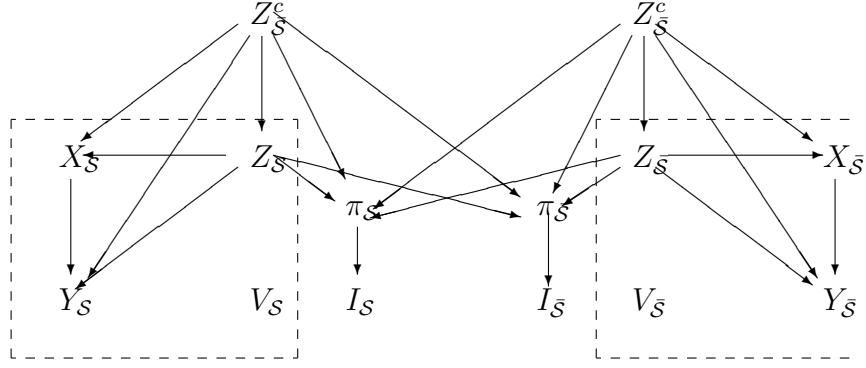

A graphical representation of the assumed conditional independencies in Assumptions $1$ and $2$ for $S=\mc{S}$ can be found in Figure \ref{fig:graph}. 

The assumption that the set of joint selection probabilities
$\{\pi_{S}:S\subseteq\mc{P}\}$ contains all information about the
sampling mechanism is natural and facilitates analysis.  In
sample surveys, the probability of selecting an observation
becomes unequal due to clustering, stratification,
post-stratification, attrition, purposive ``oversampling'' and
other non-response adjustments.  In most cases, the published
data does not contain all the design variables, thus the actual
design procedure cannot be determined.  Further, in many cases
large datasets are constructed by merging several available
datasets obtained from different surveys (e.g.
\citet{rendall_et_al_2008, tighe_livert_saxe_2010}).
Typically, each survey is based on different designs dependent on different variables.  A design for the merged dataset may not be easy to specify, but weights from individual surveys can be used to provide information about the underlying designs.        

Once the function $\pi$ is specified by the design, the first order probability of selection for $\{i\}$, $i\in\mc{P}$, is given by $\pi_i=\pi\left(\{i\},D_{\mc{P}}\right)$. The second order probabilities are similarly determined by $\pi_{ij}=\pi\left(\{i,j\},D_{\mc{P}}\right)$.  
Higher order probabilities can be specified exactly the same way.

From the graph in Figure \ref{fig:graph}, we note that $\pi_{\mc{S}}$ and $I_{\mc{S}}$ may depend on the whole of $Y_{\mc{P}}$ and $X_{\mc{P}}$ via $Z^c_{\mc{P}}$ which are not available in $\mc{S}$. Thus even though $\pi_{\mc{S}}$ does not depend on $Y_{\mc{S}}$ and $X_{\mc{S}}$ 
directly, in \eqref{eq:joint1}, $E_{\mc{P}}\left[\xi(\pi_{\mc{S}})\mid V_{\mc{S}}\right]\ne E_{\mc{P}}\left[\xi(\pi_{\mc{S}})|Z_{\mc{S}}\right]$ in general.



   

\subsection{A composite likelihood for unequal probability sampling}\label{sec:wtdemp}


We assume that the $i$th element in $\mc{S}$ is drawn with probability $\pi_i$ (i.e. $\pi_{\{i\}}$), $i=1,2,\ldots,n,$ with implied weight $d_i=\pi_i^{-1}/\sum^n_{i=1}\pi_i^{-1}$ (so that $\sum^n_{i=1}d_i=1$).  
We also assume that $\pi_i$ is positive for $i=1,2,\ldots,N$.

We consider the implication of \eqref{eq:joint1} on each $V_i$ (i.e. $V_{\{i\}}$), $i=1,2,\ldots,n$ selected in the sample.  Let $F^{(i)}_{\mc{S}}$ be the conditional distribution of $V_i$ given $\{i\}\subseteq \mc{S}$, with density $d\pf^{(i)}$.  Using Bayes' rule \citep{pfeffermann_1998}, \eqref{eq:pis} and \eqref{eq:joint1} it follows that:  
\begin{equation}\label{eq:main}
d\pf^{(i)}=\frac{Pr_{\mc{P}}(I_{\{i\}}=1,V_i)}{Pr_{\mc{P}}(I_{\{i\}}=1)}=\frac{E_{\mc{P}}\left[\pi_i\mid V_i\right]dF^0(V_i)}{Pr_{\mc{P}}(I_{\{i\}}=1)},
\end{equation}

where
\begin{equation}\label{eq:D}
Pr_{\mc{P}}(I_{\{i\}}=1)=\int Pr_{\mc{P}}(I_{\{i\}}=1,V_i)dV_i=\int E_{\mc{P}}\left[\pi_i\mid V_i\right]dF^0(V_i)dV_i. 
\end{equation}  

We call the conditional inclusion probability
 $\bp_i{\equiv}E_{\mc{P}}\left[\pi_i\mid V_i\right]$ the {\sl conditional visibility} for the $i$th element in the population and
$\Bp_i\equiv\int \bp_i dF^0(V_i)dV_i = E_{\mc{P}}[\pi_i] = E_{F^0(V_i)}[\bp_i]$
the {\sl visibility factor} for the $i$th element in the population \citep*{rao_patil_1978}.  By substituting these expressions into \eqref{eq:main} we obtain: 
\begin{equation}\label{eq:likhood}
d\pf^{(i)}=\frac{\bp_idF^0(V_i)}{\Bp_i}.
\end{equation} 
%
%
To specify $d\pf^{(i)}$ in \eqref{eq:likhood} it is typically necessary 
to model the conditional visibility ($E_{\mc{P}}\left[\pi_i\mid V_i\right]$) 
and the distribution of $V_i$ in the population ($dF^0(V_i)$).
Both of these models may depend on unknown parameters.  We denote the parameter for $dF^{0}(V_i)$ by $\theta$ and that for the model for $E_{\mc{P}}\left[\pi_i\mid V_i\right]$ by $\alpha.$

The composite likelihood for $\alpha$ and $\theta$, using all $V_i$, $i=1,2,\ldots,s$ can now be constructed as: 
\begin{equation}\label{eq:condlik}
L\left(V,\alpha,\theta\right)=\prod^n_{i=1}d\pf^{(i)}.
\end{equation} 
It is similar to the sample likelihood of \citet{pfeffermann_sverchkov_2003}.
Note that \eqref{eq:condlik} does not capture the dependence structure of the $\pf^{(i)}$.  It is a conditional likelihood if the units are drawn independently
of each other, for example, via Poisson sampling.
However, \citet*{pfeffermann_1998} show that for several designs, and under
fairly general conditions, the sampled observations in the conditional
distribution are asymptotically independent as the population 
size $N\to\infty.$ These results suggest that the  \eqref{eq:condlik}
may be a useful surrogate for the conditional likelihood in these
settings.



Notice that, \eqref{eq:condlik} is invariant to the scale of $\pi$ and $\bp$,
which can be specified up to an arbitrary positive scaling constant.
We can estimate $\bp_i$ directly from the data if the conditional distribution of $\pi_i$ given $V_i$ in $\mc{S}$ is equal to that in $\mc{P}$. Otherwise, from \citet{pfeffermann_sverchkov_1999}, we obtain $Pr_{\mc{S}}\left(\pi^{-1}_i\mid V_i\right)=\{\pi_i Pr_{\mc{P}}\left(\pi^{-1}_i\mid V_i\right)\}/E_{\mc{P}}\left[\pi_i\mid V_i\right]$. This implies:
\begin{align}
E_{\mc{P}}\left[\pi_i\mid V_i\right]&=\left[E_{\mc{S}}\left(\pi^{-1}_i\mid V_i\right)\right]^{-1}.\label{eq:expi}
\end{align}   

  


\citet{pfeffermann_1998} discuss a class of conjugate parametric models for the distribution of $V_i$
and conditional distribution with $\pi_i$ given $V_i$ such that $d\pf^{(i)}$ is in the same class as $dF^0(V_i)$. This avoids a complicated computation of $\Bp_i$.  However, estimation of $\theta$ is typically complex. The parameters in $d\pf^{(i)}$ usually depends on both $\theta$ and $\alpha$. Usually though, estimates of $\alpha$ are not of primary interest. 

Typically, $\bp_i$ would only depend on a subset of variables in $V$ which may be quite different from $A$. In particular, if the sample $\mc{S}$ was obtained by merging several subsamples drawn from different designs,  $\bp_i$ depend on the particular sample the $i$th observation belongs to. 
  Such sample indicator variables usually would not be useful in modeling the response. 


  

Parametric estimation of $F^0$ by maximizing \eqref{eq:condlik} has been discussed in \citet{rao_patil_1978}.  \citet{vardi_1985, gill_vardi_wellner_1988} consider the corresponding non-parametric likelihood when $\bp=\pi$ and study 
the empirical distribution for biased sampling models in one dimension.  

\section{Empirical likelihood to incorporate sampling weights and population level information in parameter estimation}\label{sec:emp} 



If $F_0$ is specified by a parametric family $\mc{F}_{\theta}$, a natural way to include population level information may be to maximize \eqref{eq:condlik} under the constraint that, $C(\theta)=E_{F_{\theta}}\left[g\left(V_1\right)\right]=\int g\left(V_1\right)dF_{\theta}=\gamma$. 
However, since $C(\theta)$ is usually a non-linear function of $\theta$, such maximization is computationally difficult and becomes almost infeasible even for a moderate number of covariates and population level constraints. Furthermore, when using \eqref{eq:condlik}, analytical expressions are available only if one restricts to conjugate families of distributions.  
 Outside this class $\Bp_i$ has to be computed numerically (see \citet{pfeffermann_sverchkov_2003}) which may be time consuming. Correct specification of the joint distribution of $V_i$ is difficult in many situations. 
   
An alternative is to use empirical likelihood \citep{owenbook} and estimate $F^0$ non-parametrically from the observed weighted sample and include all the available parametric or population based information in the analysis.
 



Suppose that, for each $F\in\mc{F}$, $w_i=F(\left\{V_i\right\})$
be the weights $F$ assigns on $V_i$ ($w_i=0$ for all $F$ continuous at $V_i$).  Let $\Delta_{n-1}$ denote the $n$ dimensional simplex,
\begin{align}
\mc{W}_{\theta}&=\left\{w\in\Delta_{n-1}~:~\sum^n_{i=1}w_i\psi_{\theta}\left(Y_i,A_i\right)=0\right\}~~~~~~~~~\text{for each $\theta\in\Theta$,}\label{eq:thetaset}\\
\mc{W}_P&=\left\{w\in\Delta_{n-1}~:~\sum^n_{i=1}w_ig\left(V_i\right)=\gamma\right\}\text{  and }
\mc{W}=\bigcup_{\theta\in\Theta}\left(\mc{W}_{\theta}\cap\mc{W}_P\right).\label{eq:wset}
\end{align}





\noindent{\bf Assumption $3$: Label-independence of the visibility factors.} From Assumption 1, 
$\Bp_i=E_{\mc{P}}[\pi\left(i,D_{\mc{P}}\right)].$ If these do not depend on the population
labels but only on the design variables,
$\Bp_i=\Bp\left(D_{\mc{P}}\right)\equiv\Bp$ \citep*{godambe75,hartleyrao75}. Hence, 
each element in the population and sample will have equal visibility factor.




Under Assumptions 1, 2 and 3, the natural empirical composite log-likelihood function corresponding to \eqref{eq:condlik} is obtained by substituting $dF^{0}_i=F^{0}(\{V_i\})$ by $w_i$ and $\Bp$ by $\hat{\Bp}=\sum^n_{i=1}\bp_iw_i$. It takes the form:
\begin{equation}\label{eq:logemplik}
L_{CE}(w,\bp)=\sum^n_{i=1}\log(w_i)-n\log(\sum^n_{i=1}\bp_iw_i).
\end{equation}
 
In presence of parametric and population level information we estimate the weights $\h{w}_{CE}$ as $\argmax_{w\in\mc{W}}L_{CE}(w,\bp)$.
A constrained estimator $\h{\theta}_{CE}\in\Theta$ can be obtained as \citep{Qinlawless1,chaudhuri_handcock_rendall_2008} 
\begin{equation}\label{eq:theta}
\h{\theta}_{CE}=\argmax_{{\kern-20pt}\theta\in\Theta}\left\{\max_{w\in\mc{W}_{\theta}\cap\mc{W}_P}\left(L_{CE}(w,\bp)\right)\right\}.
\end{equation}



\citet{kim_2009} considers estimation of population mean under Poisson sampling and uses expression \eqref{eq:logemplik} with $\bp_i$ replaced by $\pi_{i}$.  In the context of two sample testing, \citet{qin_1993} maximizes \eqref{eq:logemplik} w.r.t. 
$w_{i}$ and $\Bp$ with the additional constraint $\sum^n_{i=1}w_i\pi_i=\Bp$.  Similar approaches have been taken by \citet{qin_leung_shao_2002, qin_zhang_2007}
 to include auxiliary information in the presence of non-ignorable data.

Choice of $\hat{\Bp}$ in the second term of \eqref{eq:logemplik} is crucial.  Our choice involves both $\bp_{i}$ and $w_{i}$.  Use of the sample mean of $\pi$ or $\bp$ would lead to unweighted estimator of the parameters.

We follow \citet{pfeffermann_sverchkov_1999,pfeffermann_sverchkov_2003} and estimate $\alpha$ separately from $w$.  In particular, $\h{\alpha}$, the maximum likelihood estimator for $\alpha$
obtained under the model for $E_{\mc{P}}[\pi|V]$ is used to obtain $\bp$. In most cases, our main interest is in finding $\bp$, not $\hat{\alpha}$.  



 
\begin{thm}\label{thm:FCE}
Suppose $\mc{W}_{P}=\Delta_{n-1}$.  The estimate of $F_0$ obtained by maximizing \eqref{eq:logemplik} above over $\mc{W}_{P}$ is given by:
\begin{equation}
\h{F}_{CE}(C)=\sum^n_{i=1}\frac{(1/\bp_i)}{\sum^n_{i=1}(1/\bp_i)}{\bf 1}_{\{V_i\in C\subseteq \mbb{R}^{m+p+1}\}}.
\end{equation} 
\begin{proof} See Appendix.\end{proof}
\end{thm}

Thus $\h{w}_{CE}$ has the desirable property that when $\mc{W}_{\mc{P}}=\Delta_{n-1}$, the first step of the two-step method gives $\h{w}_{CEi}=\bp^{-1}_i/\sum^n_{i=1}\bp^{-1}_i$, for all $i$. In this situation, $\h{\theta}_{CE}$ satisfies:
\begin{equation}\label{eq:JR}
\sum^n_{i=1}\frac{1}{\bp_i}\psi_{\theta}\left(y_i,a_i\right)=0
\end{equation}

Estimators based on \emph{inverse probability weighted} score functions, as in \eqref{eq:JR} have been studied in details in the statistics literature. They occur very often in connection with missing data, two-phase designs, etc.  However, since the weights $\bp$ are random, their justification as an usual Horvitz-Thompson type estimator is not entirely appropriate.  
\citet{beaumont_2008} regards this as a smoothed Horvitz-Thompson estimator.  Our framework avoids invoking Horvitz-Thompson estimators, and provides a better explanation in terms of conditional empirical likelihood.  Furthermore, the derivation follows naturally from a likelihood framework. 
The resulting log-likelihood is also different from a typical weighted log-likelihood found in the literature.
This can be exploited in Bayesian formulations of related problems specially in small-area estimation and in multi-phase sampling sampling where the design in the later phases depend on the observed variables in the earlier phase \citep{breslow_wellner_2006}.
 


\section{Computational methods for parameter estimates}\label{sec:comp}
 Since $\mc{W}$ is specified by linear constraints on weights, empirical likelihood based methodology has a clear computational advantage over the corresponding constrained maximum likelihood estimator.  This computational burden can further be reduced by using a two-step procedure described in \citet{chaudhuri_handcock_rendall_2008}. 

  
In this adaption, the first maximization is done over $\mc{W}_P$. These maximizing weights are then substituted in the estimating equation for $\theta$ and in the second step these equations are solved to obtain the parameter estimates using standard Newton-Raphson method.
  
Since the log-empirical likelihood in \eqref{eq:logemplik} is concave on a closed convex set $\mc{W}_P$, it has a unique maximize on $\mc{W}_P$. 
Clearly if this maximizing weights are in $\mc{W}_{\theta^{\prime}}$ for some $\theta^{\prime}\in\Theta$, $\h{\theta}=\theta^{\prime}$.  The two-step method may fail for a situation
where there is a small sample size and when a solution to the second step does not exist. Such situations are rare in practice.


Maximizing \eqref{eq:logemplik} over $\mc{W}_P$ requires some discussion.  The objective function is:
\begin{equation*}
L\left(w,\lambda_1,\lambda_2\right)=\sum^n_{i=1}\log(w_i)-n\log(\sum^n_{i=1}\bp_iw_i)-\lambda_1\left(\sum^n_{i=1}w_i-1\right)-n\lambda_2\sum^n_{i=1}w_ih_i,
\end{equation*}
where $\lambda_1$ and $\lambda_2$ are Lagrange multipliers and $h_i=g\left(Y_i,A_i\right)-\gamma$.

By differentiating w.r.t. $w_i$ and following \citet{owenbook} mutatis mutandis, we obtain $\lambda_1=0$.  So by writing $\kappa=\lambda_2\sum^n_{i=1}\bp_iw_i$ we obtain (similar to \citet{kim_2009}) 
\begin{equation}\label{eq:whatcs}
w_i=\frac{\sum^n_{i=1}\bp_iw_i}{n}\frac{1}{\bp_i+\kappa h_i}.
\end{equation} 

Clearly $w_i\le 1$, for each $i$ implies the restriction

\begin{equation}\label{eq:restr}
n\left\{\bp_i+\kappa h_i\right\}\ge\sum^n_{i=1}\bp_iw_i=\Bp.
\end{equation}

By substituting these values of $w_i$ into \eqref{eq:logemplik} we obtain
\begin{equation}\label{eq:kappa}
L_{CE}(w,\bp)=-\sum^n_{i=1}\log\left(\bp_i+\kappa h_i\right)-n\log(n).
\end{equation}

The weights can be estimated by minimizing $L_{CE}(w,\bp)$ w.r.t. $\kappa$ under the restriction in \eqref{eq:restr} for all $i=1,2,\ldots,n$.

From \eqref{eq:restr} it is clear that the lower bound on $\left(\bp_i+\kappa h_i\right)$ depends on the weights which are unknown and thus direct constrained minimization of $L_{CE}(w,\bp)$ does not follow from \citet{owenbook} in a straightforward manner.  
However, The following Lemma shows that $\h{w}_{CEi}$ can be obtained, by solving an easier but similar optimization problem: 
  



\begin{lem}\label{lem:reint} 
Suppose $w^{\star}=\argmax_{w}\sum^n_{i=1}\log w_i$ subject to
$w\in\Delta_{n-1}$ and $\sum^n_{i=1}w_i(h_i/\bp_i)=0$.  Then $\h{w}_{CEi}=(w^{\star}_i/\bp_i)/\sum^n_{i=1}(w^{\star}_i/\bp_i)$.



\begin{proof} See Appendix.\end{proof}
\end{lem}


We show later (see \eqref{eq:hwstar}) that $\h{w}^{\star}=\bp_i\h{w}_{CEi}/\h{\Bp}$.  Thus $\h{w}^{\star}$ correspond to constrained empirical likelihood estimator of $\pf$. Parameter constraints can also be included.  One computes 
\begin{equation}\label{eq:esthstar}
\h{w}^{\star}=\argmax_{{\kern-20pt}w}\sum^n_{i=1}\log(w_i)
\end{equation}
\centerline{subject to $w\in\Delta_{n-1}$, $\sum^n_{i=1}w_i(\psi_{\theta}\left(Y_i,A_i\right)/\bp_i)=0$ and $\sum^n_{i=1}w_i(h_i/\bp_i)=0$.}\\

Notice that, the constrained problem in \eqref{eq:esthstar} is analogous to \citet{owenbook} for which standard software are available (e.g. \citet{chen_sitter_wu_2002}). 






\section{Two alternative estimators}
\subsection{The pseudo-maximum likelihood estimator}
Suppose that $F_0$ belongs to the parametric family $\mc{F}_{\theta}$ and that $S_{\theta}$ is the corresponding score function.  Then the \emph{pseudo-maximum likelihood estimator} \citep{krieger_pfeffermann_1992} of $\theta$ (i.e. $\h{\theta}_{PL}$) with no constraints is obtained by solving:
\begin{equation}\label{eq:score}
\sum^n_{i=1}d_iS_{\theta}\left(Y_i,A_i\right)=0.
\end{equation} 

Under standard assumptions, $\h{\theta}_{PL}$ is asymptotically unbiased and normally distributed.  Its asymptotic variance can also be computed analytically \citep{chambers_2003}. Unfortunately, there is no clear way to incorporate population level information in $\h{\theta}_{PL}$.  In what follows, we compare the efficiency of this estimator to other estimators. 
It is expected to be at a disadvantage as it does not incorporate any population level information.  For similar unconstrained parametric estimates we refer to \citet{pfeffermann_sverchkov_2003}.

\subsection{The pseudo-empirical likelihood estimator} 

Motivated by the idea that super-population parameters would be closely
approximated by their large finite-population counterparts, 
\citet{chen_sitter_1999} introduced a \emph{pseudo-empirical likelihood estimator} (PELE) to include fixed sampling weights $d_i$. They estimate the total of $\log(w_i)$ in the population through the design unbiased Horvitz-Thompson estimator from the sample.  Their estimator for $w$ is given by:
\begin{equation}\label{eq:hwCS}
\h{w}_{CS}=\argmax_{{\kern-20pt}w\in\mc{W}}\sum^n_{i=1}d_i\log w_i.
\end{equation}

This estimator has been frequently used in sampling literature in several contexts. \citet{rao_wu_2008} interpret of \eqref{eq:hwCS} as a ``backward'' Kullback-Leibler divergence between $w$ and $d$. $\h{w}_{CS}$ in \eqref{eq:hwCS} minimises this divergence.  Parameter estimates $\h{\theta}_{CS}$ can be obtained similarly as in \eqref{eq:theta}.  
It is known that for the population mean, under certain conditions PELE is asymptotically equivalent to the generalized regression (GREG) estimator.  
For stratified single stage and multi-stage sampling PELE is equivalent to the optimal regression estimator (ORE), but in many other cases PELE may be substantially better than the ORE.  Further discussion on this estimate may be found in \citet{glenn_zhao_2007,wu_rao_2006,fu_wang_wu_2008} among others. 
In the absence of any population level information $\h{\theta}_{CS}$ reduces to $\h{\theta}_{PL}$ described above.

  
 
The two-step estimation method developed for $\h{\theta}_{CE}$ can be adapted to 
$\h{\theta}_{CS}$. However they are fundamentally different and lead to a completely different profile likelihoods for $\theta$.

 


\section{Asymptotic properties of the estimators}\label{sec:asymp}

In this section we discuss the asymptotic properties of the two parameter estimates of $\theta$ obtained from the two empirical likelihood based methods (CE and CS) under the true population distribution $F^0$ as $N\rightarrow\infty$. Only two-step estimation of $\h{\theta}_{CS}$ and $\h{\theta}_{CE}$ are considered and compared with $\hat{\theta}_{PL}$. 
In the last part of this section we discuss how to estimate their asymptotic standard errors form the sample.  For a formal setup of $N\rightarrow\infty$ we refer to \citet{fuller_2009}.

We first discuss notation and specify the assumptions.  Let us denote :
\begin{align}
f_1\left(v,d,\theta,\lambda\right)&=\frac{d}{1+\lambda h\left(v,\gamma\right)}\left(\psi_{\theta}(y,a),h\left(v,\gamma\right)\right)\\
f_2\left(v,\bp,\theta,\kappa\right)&=\frac{1}{\bp+\kappa h\left(v,\gamma\right)}\left(\psi_{\theta}(y,a),h\left(v,\gamma\right)\right)
\end{align}

Suppose $\theta_0$ is the true value of $\theta$.  We make the following assumptions.
\begin{list}{}{}
\item[A.$1.$] We assume that under $F_0$ both $f_1\left(v_i,d_i,\theta,\lambda\right)$, $1\le i\le N$ and $f_2\left(v_i,\bp_i,\theta,\kappa\right)$, $1\le i\le N$ are i.i.d. random vectors for all $\theta$ and $\lambda$.  
\item[A.$2.$] Suppose for all $d$ and $\bp$, $E_{\mc{P}}\left[\fd\right]=E_{\mc{P}}\left[\fp\right]=0$.
\item[A.$3.$] The rest of the regularity conditions are standard and similar to \citet{Qinlawless1} and \citet{serfling1}. Details can be found in \citep{blinded}. 
\end{list}

Assumption A.$1$ is not the most general possible, but it is sufficient to illustrate the asymptotic properties of our estimator.  This assumption is not too restrictive. Note that in the population by assumption $V_i$, $1\le i\le N$ are i.i.d.. So if $d_i$ and $v_i$ are independent and identically distributed, A.$1.$ would hold.  
As for example, $f_2\left(v_i,\bp_i,\theta,\kappa\right)$ are i.i.d. if $\pi_i=\pi\left(\{i\},D_{i}\right)$, i.e. $\pi_i$ only depends on the $i$th observation in $\mc{P}$. In this case, $d_i$ would be identically distributed but weakly dependent ($\sum^N_{i=1}d_i=1$ by assumption).  
Even if $\pi_i$ and $\pi_j$ are not independent, in many cases, judicious choice of variables in $V$ can make $\bp_i$ and $\bp_j$ independent for all $i\ne j$.  Furthermore, under certain assumption, similar to \citet{pfeffermann_1998} we can show that for large $N$, 
the composite likelihood in \eqref{eq:condlik} is very close to the  composite likelihood of the whole sample.      
  



Let us denote
\begin{align}
\psi^{\prime}\left(y,a,\theta\right)&=\partial\psi(y,a,\theta)/\partial\theta,
G=E_{\mc{P}}\left[d_1\psi^{\prime}\left(y_1,a_1,\theta_0\right)\right],
G^{\star}=E_{\mc{P}}\left[d^2_1\psi^2\left(y_1,a_1,\theta_0\right)\right],\nonumber\\
K_1&=E_{\mc{P}}\left[d_1\psi\left(y_1,a_1,\theta_0\right)h(v_1,\gamma)\right],
K_2=E_{\mc{P}}\left[d^2_1\psi\left(y_1,a_1,\theta_0\right)h(v_1,\gamma)\right],\nonumber\\
H_1&=E_{\mc{P}}\left[d_1h^2(v_1,\gamma)\right],
H_2=E_{\mc{P}}\left[d^2_1h^2(v_1,\gamma)\right],\nonumber\\
\mG&=E_{\mc{P}}\left[\psi^{\prime}\left(y_1,a_1,\theta_0\right)/\bp_1\right],
\mG^{\star}=E_{\mc{P}}\left[\psi^2\left(y_1,a_1,\theta_0\right)/\bp^2_1\right],\nonumber\\
\mK_2&=E_{\mc{P}}\left[\psi\left(y_1,a_1,\theta_0\right)h(v_1,\gamma)/\bp^2_1\right],
\mH_2=E_{\mc{P}}\left[h^2(v_1,\gamma)/\bp^2_1\right].\nonumber
\end{align}

The next two theorems prove the strong consistency and the asymptotic normality of $\h{\theta}^{(N)}_{CS}$ and $\h{\theta}^{(N)}_{CE}$ respectively. The proofs can be found in the supplement \citep{blinded}. 

\begin{thm}\label{thm:CS}
Under the assumptions A.$1.$ - A.$3.$ almost surely the equation $\sum^N_{i=1}f_1\left(v_i,d_i,\theta,\lambda\right)=0$ admits a sequence of solutions $(\h{\theta}^{(N)}_{CS},\h{\lambda}^{(N)})$ such that  

\begin{enumerate}{}{}
\item $(\h{\theta}^{(N)}_{CS},\h{\lambda}^{(N)})$ $\longrightarrow$ $(\theta_0,0)$ almost everywhere as $N\rightarrow\infty$,
\item $N^{1/2}(\h{\theta}^{(N)}_{CS}-\theta_0)\Rightarrow N\left(0,\mathcal{V}_{CS}\right)$ distribution as $N\rightarrow\infty$, where\\
$\mathcal{V}_{CS}=G^{-1}\left(G^{\star}-K_1H^{-1}_1K^T_2-K_2H^{-1}_1K^T_1+K_1H^{-1}_1H_2H^{-1}_1K^T_1\right)\left(G^T\right)^{-1}$,
\item $N^{1/2}\h{\lambda}^{(N)}\Rightarrow N\left(0,H^{-1}_1H_2H^{-1}_1\right)$ distribution as $N\rightarrow\infty$,
\item Asymptotic covariance of $\h{\theta}^{(N)}_{CS}$ and $\h{\lambda}^{(N)}$ is given by $G^{-1}\left(K_1H^{-1}_1H_2-K_2\right)H^{-1}_1$.
\end{enumerate}
\end{thm}
 
\begin{thm}\label{thm:CE}
Under the assumptions A.$1.$ - A.$3.$ almost surely the equation $\sum^N_{i=1}f_2\left(v_i,\bp_i,\theta,\kappa\right)=0$ admits a sequence of solutions $(\h{\theta}^{(N)}_{CE},\h{\kappa}^{(N)})$ such that  

\begin{enumerate}{}{}
\item $(\h{\theta}^{(N)}_{CE},\h{\kappa}^{(N)})$ $\longrightarrow$ $(\theta_0,0)$ almost everywhere as $N\rightarrow\infty$,
\item $N^{1/2}(\h{\theta}^{(N)}_{CE}-\theta_0)\Rightarrow N\left(0,\mathcal{V}_{CE}\right)$ distribution as $N\rightarrow\infty$, where \\$\mathcal{V}_{CE}=\mG^{-1}\left(\mG^{\star}-\mK_2\mH^{-1}_2\mK^T_2\right)\left(\mG^T\right)^{-1}$,
\item $N^{1/2}\h{\kappa}^{(N)}\Rightarrow N\left(0,\mH^{-1}_2\right)$ distribution as $N\rightarrow\infty$,
\item $\h{\theta}^{(N)}_{CE}$ and $\h{\kappa}^{(N)}$ are asymptotically independent as $N\rightarrow\infty$.
\end{enumerate}
\end{thm} 


Suppose $\h{\theta}_{PL}$ is the unconstrained pseudo maximum likelihood estimator. It can be shown that the variance-covariance matrix of $\sqrt{N}\left(\h{\theta}^{(N)}_{PL}-\theta_0\right)$ is given by $G^{-1}G^{\star}\left(G^T\right)^{-1}$. 
$\h{\theta}_{CS}$ and $\h{\theta}_{CE}$ are both constrained by population level information. Thus it is natural to expect that these two estimator would be more efficient than $\h{\theta}_{PL}$. Theorem \ref{thm:CS} does not ensure any reduction in the standard error of $\h{\theta}_{CS}$. 
$\h{\theta}_{PL}$ does not use the same weights as $\h{\theta}_{CE}$, so no conclusion can be drawn based on Theorem \ref{thm:CE}.  However, in the majority of cases $\h{\theta}_{CE}$ 
is more efficient than $\h{\theta}_{CS}$ and $\h{\theta}_{PL}$.  

Note that, in Theorem \ref{thm:CS}, $\h{\theta}^{(N)}_{CS}$ and
$\h{\lambda}^{(N)}$ are not asymptotically independent, and this heuristically
explains why the asymptotic variance of $\theta^{(N)}_{CS}$ does not always
decrease with more constraints.
In fact, we show that if $\h{\theta}^{(N)}_{CS}$ and $\h{\lambda}^{(N)}$ are
asymptotically independent $\h{\theta}^{(N)}_{CS}$ would be at least as
efficient as $\h{\theta}^{(N)}_{PL}$:

\begin{cor}\label{cor:CS} 
Under the conditions of Theorem \ref{thm:CS} if $\h{\theta}^{(N)}_{CS}$ and $\h{\lambda}^{(N)}$ are asymptotically independent, then $\mathcal{V}_{CS}=G^{-1}\left(G^{\star}-K_2H^{-1}_2K^T_2\right)\left(G^T\right)^{-1}$.
\end{cor}
One situation (under A.$1$ to A.$3$) where the conditions of Corollary \ref{cor:CS} hold is when $d_1$ is independent of $\psi\left(y_1,a_1,\theta_0\right)$, $h\left(v_1,\gamma\right)$.  

\subsection{The case where the conditional and unconditional visabilities are equal}

We know consider the important special case where $\bp=\pi$.  In this case, clearly, all of $\h{\theta}_{PL}$, $\h{\theta}_{CS}$ and $\h{\theta}_{CE}$ depend on the same weights.  Thus, if the assumptions A.$1.$ to A.$3.$ hold, from Theorem \ref{thm:CE} it follows that $\h{\theta}_{PL}$ is less efficient than $\h{\theta}_{CE}$.  In fact, in the following theorem we show that $\h{\theta}_{CE}$ is more efficient than $\h{\theta}_{CS}$ as well.

\begin{thm}\label{thm:stdgain}
If $\bp_i=\pi_i$, for all $i=1,2,\ldots,N$, then under the assumptions A$1.$ - A$3.$ the asymptotic standard error of $\h{\theta}^{(N)}_{CS}$ is at least as large as that of $\h{\theta^{(N)}}_{CE}$. Furthermore, the equality holds if in Theorem \ref{thm:CS}, $\h{\theta}^{(N)}_{CS}$ and $\h{\lambda}^{(N)}$ are asymptotically independent.
\end{thm}




If the design puts equal probability on all sampled observations, $\h{\theta}_{CS}=\h{\theta}_{CE}$.  If all design variables are observed, i.e. $\pi_i=\pi\left(\{i\},Z_i\right)$, $\h{\theta}_{CE}$ gains a lot over $\h{\theta}_{CS}$ in terms of efficiency.  In fact, even if $\bp\ne\pi$, $\h{\theta}_{CE}$ usually has lower standard error than $\h{\theta}_{CS}$.  
\citet{kim_2009} considers estimation of population mean and discusses conditions when $\h{\theta}_{CS}$ could be more efficient that $\h{\theta}_{CE}$. However, his simulation studies as well as ours (not presented here) show no major gain in efficiency for $\h{\theta}_{CS}$ in any situation, unless assumption A.$1.$ was strongly violated.


\subsection{Estimating asymptotic covariance matrices}
Theorem \ref{thm:CE} and \ref{thm:CS} are based on $N\rightarrow\infty$.
In practice, the asymptotic covariance matrices of $\h{\theta}_{CS}$ and $\h{\theta}_{CE}$ need to be estimated from the available data.  We estimate them directly from their respective expressions. In particular the estimates are given by:
\begin{align}
\h{G}&=\sum^n_{i=1}\h{w}_{CSi}d_i\psi^{\prime}\left(y_i,a_i,\h{\theta}_{CS}\right),
\h{G^{\star}}=\sum^n_{i=1}\h{w}^2_{CSi}d^2_i\psi^2\left(y_i,a_i,\h{\theta}_{CS}\right),\nonumber\\
\h{K}_1&=\sum^n_{i=1}\h{w}^2_{CSi}d_i\psi\left(y_i,a_i,\h{\theta}_{CS}\right)h(v_i,\gamma),
\h{K}_2=\sum^n_{i=1}\h{w}^2_{CSi}d^2_i\psi\left(y_i,a_i,\h{\theta}_{CS}\right)h(v_i,\gamma),\nonumber\\
\h{H}_1&=\sum^n_{i=1}\h{w}^2_{CSi}d_ih^2(v_i,\gamma),
\h{H}_2=\sum^n_{i=1}\h{w}^2_{CSi}d^2_ih^2(v_i,\gamma),\nonumber\\
\h{\mG}&=\sum^n_{i=1}\h{w}_{CEi}\psi^{\prime}\left(y_i,a_i,\h{\theta}_{CE}\right)/\bp_i,
\h{\mG}^{\star}=\sum^n_{i=1}\h{w}^2_{CEi}\psi^2\left(y_i,a_i,\h{\theta}_{CE}\right)/\bp^2_i,\nonumber\\
\h{\mK}_2&=\sum^n_{i=1}\h{w}^2_{CEi}\psi\left(y_i,a_i,\h{\theta}_{CE}\right)h(v_i,\gamma)/\bp^2_i,
\h{\mH}_2=\sum^n_{i=1}\h{w}^2_{CEi}h^2(v_i,\gamma)/\bp^2_i.\nonumber
\end{align} 

The estimated values $\h{\mathcal{V}}_{CS}$ and $\h{\mathcal{V}}_{CE}$ at $\h{\theta}_{CS}$ and $\h\theta_{CE}$ can be found by substituting the above expression in the formulas in Theorems \ref{thm:CS} and \ref{thm:CE}.

The properties of the above estimates depend on the specifics
of the sampling design $D$. 
In our experience with simulated and real data, these estimates are close to their target values and also those obtained from non-parametric bootstraps (results not shown). 

\section{Application to demographic hazard modeling with a complex longitudinal survey}\label{sec:data}
We use 1985-97 years of the Panel Study of Income Dynamics \citep[PSID,][]{PSID2010}
in combination with population-level birth registration data from the National Center for Health Statistics (NCHS) age-specific first birth probabilities in the US \citep{schoen_2005} to estimate the relationship between the probability of first birth to age and other socio-demographic factors.

\begin{figure}[t]
\begin{center}
\parbox[c][.65\columnwidth][t]{.45\columnwidth}{\subfigure[Population (NCHS) and sample survey (PSID) first birth probabilities for age $17$ to $30$, $1985-97$.\label{fig:asfbr_a}]{\rotatebox{0}{\resizebox{.5\columnwidth}{.45\columnwidth}{\includegraphics{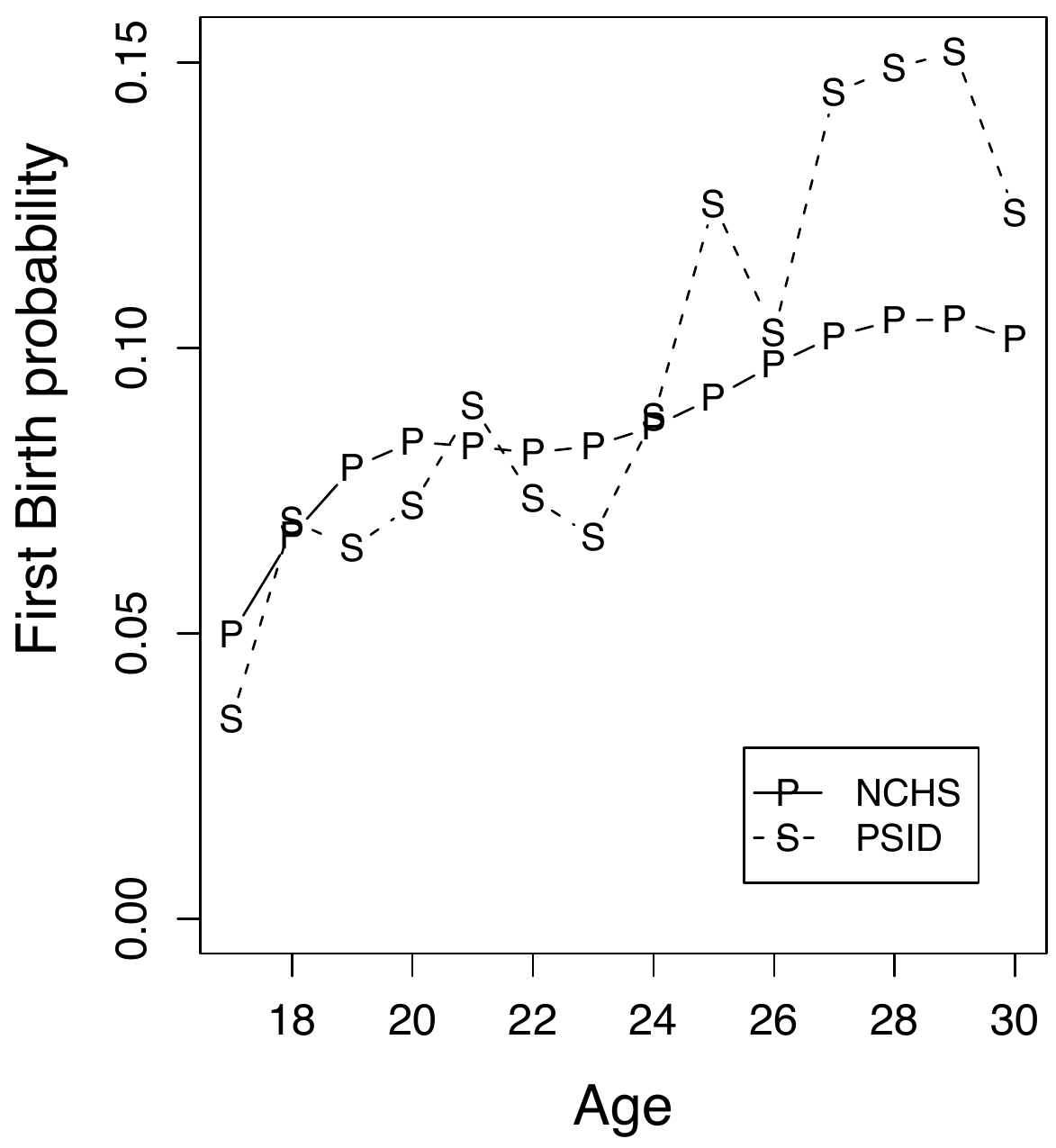}}}}}\vspace{0mm}\hspace{.05\columnwidth}
\parbox[c][.65\columnwidth][t]{.45\columnwidth}{\subfigure[t][Distribution of sample weights in two PSID subsamples.\label{fig:asfbr_b}]{\rotatebox{0}{\resizebox{.5\columnwidth}{.45\columnwidth}{\includegraphics{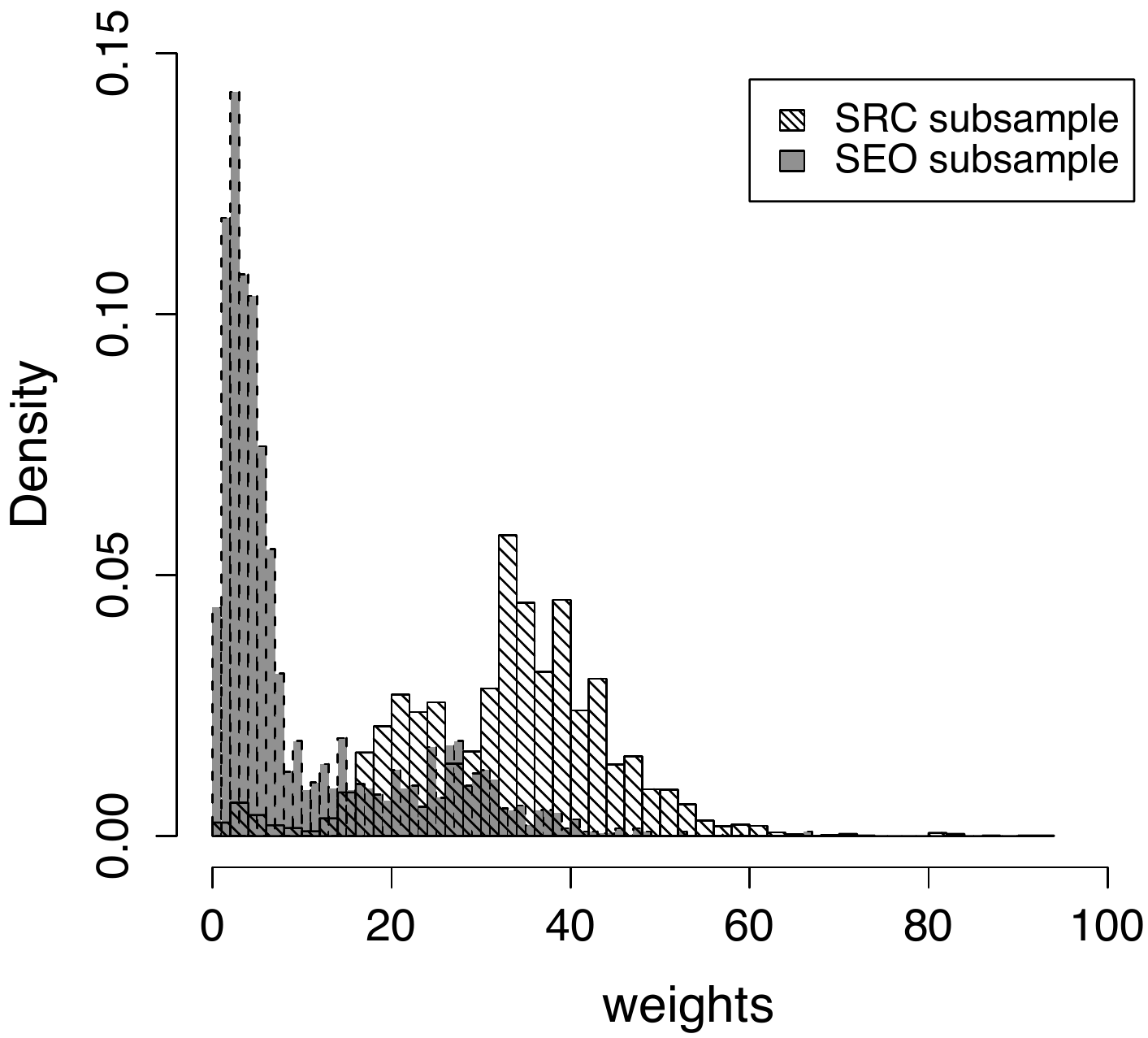}}}}}
\vskip -52pt
Figure 2
\end{center}
\end{figure}
\setcounter{figure}{2}
The PSID design incorporates a high degree of stratification on variables associated with first birth timing, and additionally incorporates clustering within geographic areas and within families.  
Most important to the implementation of our composite empirical likelihood (CE) estimator is the incorporation of the unequal probabilities of selection produced by the PSID stratified design.
This consists of two subsamples: an equal probability ``Survey Research Center'' (SRC) subsample and a low-income population oversample ``Survey of Economic Opportunity'' (SEO) subsample.
%
Thus the weights of the SEO-low-income observations (Family IDs $> 5,000$) are about one-tenth of those in the main SRC sample (Family IDs $<3,000$, see Figure \ref{fig:asfbr_b}).  However, when used with the sample weights supplied with the data, the combined subsamples are designed to be nationally representative.
A comparison of the weighted PSID sample estimates of age-specific first birth probabilities to population-level NCHS probabilities for approximately the same period is seen in \ref{fig:asfbr_a}.  The NCHS data display the humped pattern found in very large-scale sample survey data \citep{sullivan_2005}.  
The probability initially peaks at the age $20$ then drops off and then climbs again to a maximum at age $28$.  
Sampling error in the weighted PSID estimates makes it difficult to discern a pattern beyond the similarly upward trend in age-specific birth probabilities over this age range.


The sampling scheme in the PSID followed not only all original
$1968$ household members but also all of their descendants.  This
within-family design results in a highly clustered sample in two
ways: first in terms of a common family developmental environment
and genetic inheritance, and second in terms of common geographic
locational features at least while growing up and often also into
adulthood.  While the CE estimator can be applied to data with
clustering, it will be more appropriate if the level of
clustering is not extreme and we therefore take steps to remove
the family-based clustering built into the PSID design.
Given family clusters of size $n_f$ for the $f$th family, we
construct $10000$ de-clustered datasets, by randomly choosing one
member from each family with probability $1/n_f$.  The PSID
weights of the selected units are adjusted for this de-clustering
by multiplying them by their corresponding $n_f$.
There is also geographic clustering of the PSID sample across
families within the primary sampling units from which the
original $1968$ families were selected, but we ignore this
additional element of clustering.

The probability of giving the first birth in year $t-1$ and $t$ is modeled as:

\begin{align}\label{eq:model}
\mbox{log-odds}(I^{(B)}_i)=\theta_0+\sum_{l\in C}\theta_{l}I^{(l)}_i+\sum^{30}_{k=18}\theta_kI^{(k)}_i+\theta_{MD}\cdot MD_i+\theta_{MD^2}\cdot MD^2_i.
\end{align}


Here $I^{(l)}_i=1$ and $I^{(k)}_i=1$ if observation $i$ is in category $l$ and $k$ respectively.  The variables $k$ denotes the age at time $t$, $k=18,\ldots,30$ and $B$ denotes the first birth in the year $t-1$ to $t$.  The set $C$ consists of the following variables: $M$=married at $t-1$, $PM$= previously married at $t-1$, 
$E$= employed at $t-1$, $W$= white, $HS$= high school graduate at
$t-1$, $NHNS$= non HS and not in school at $t-1$, $NHS$= non HS
and in school at $t-1$, $COLL$= any college education at $t-1$.
The variable $MD$= marital duration at $t-1$.

The weight ($d_i$) of an observation depends strongly on 
the sample (i.e. SRC/SEO) it comes from. Suppose $I^{(S)}_i=1$ if the observation is from the SRC sample.  
We assume that $A$ is the set of all variables appearing in \eqref{eq:model} and  $V=A\cup I^{(S)}$. 
   
It is not clear if the distribution of $\pi\mid V_i$ in the sample is same as in the population.  
Thus we use \eqref{eq:expi} to estimate $\bp$.  We take $E_{\mc{S}}\left[d_i\mid V_i\right]$ to be the fitted values from a Gamma regression with inverse (canonical) link function of $d_i/n_f$ on $V_i$, multiplied by $n_f$.  The model for the mean function  was taken to be:  
\begin{equation}\label{eq:wt} 
\alpha_0+\sum^{30}_{k=20}\alpha_k I^{(k)}_i+\alpha_W I^{(W)}_i + \alpha_S I^{(S)}_i +\alpha_{SW} (I^{(W)}_{i}\cdot I^{(S)}_{i}).
\end{equation}


We use NCHS values of the age specific first birth probability given in \citet{schoen_2005} as our population level constraints. For $k=17,18,\ldots,30$ the constraints are given by:
\begin{equation}
\sum^n_{i=1}w_iI^{(k)}_i\left(I^{(B)}_i-\gamma_k\right)=0.
\end{equation}

Following \citet{mshren2}, we expect only the 
coefficients $k$ for ages (plus the intercept representing the reference age) to have their standard errors substantially reduced by the introduction of population information
in the constraint, even while all standard errors will be at least as low as for the unconstrained model.

\begin{figure}[t]
\begin{center}
\resizebox{.6\linewidth}{.4\linewidth}{\includegraphics{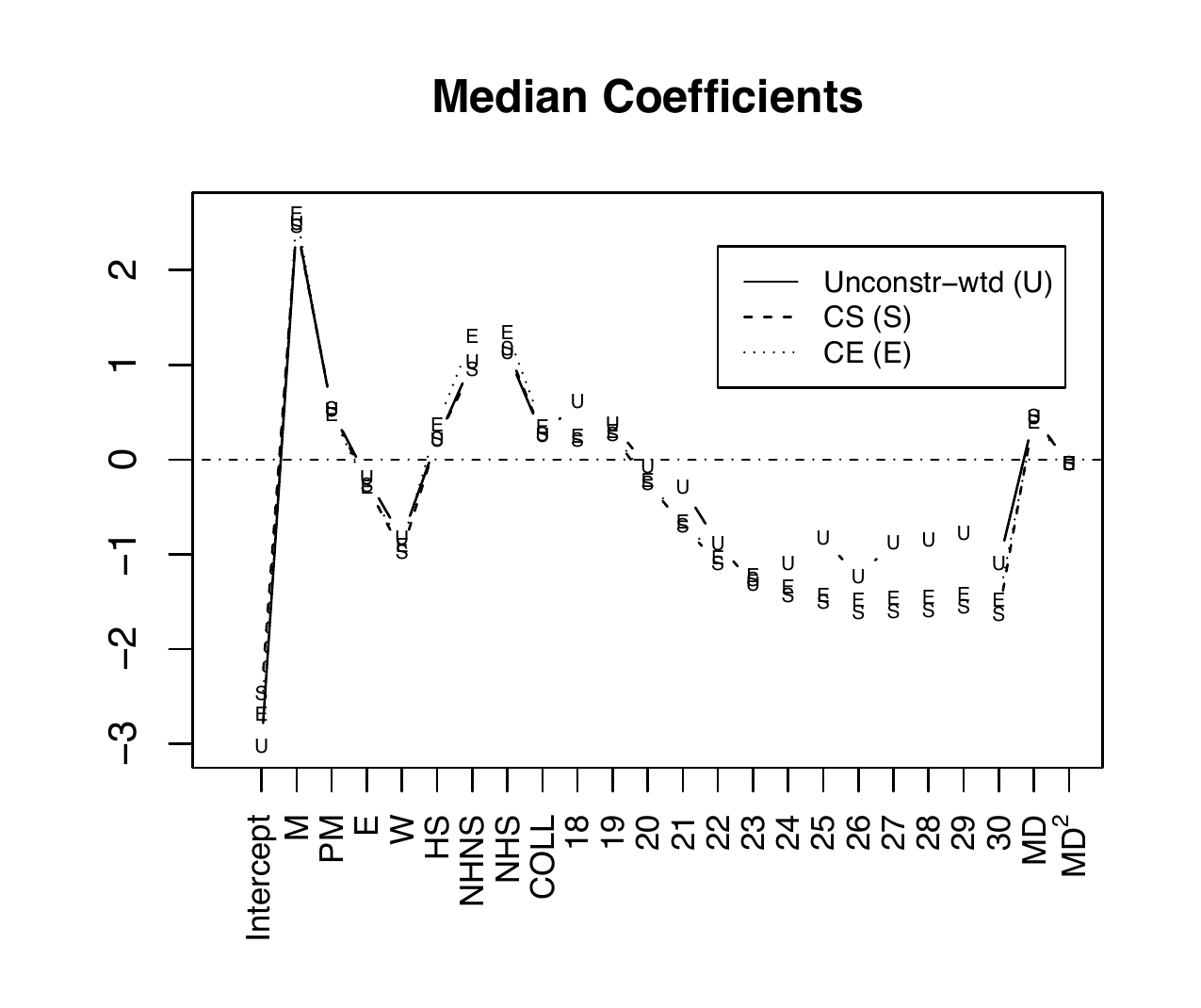}}
\resizebox{.6\linewidth}{.4\linewidth}{\includegraphics{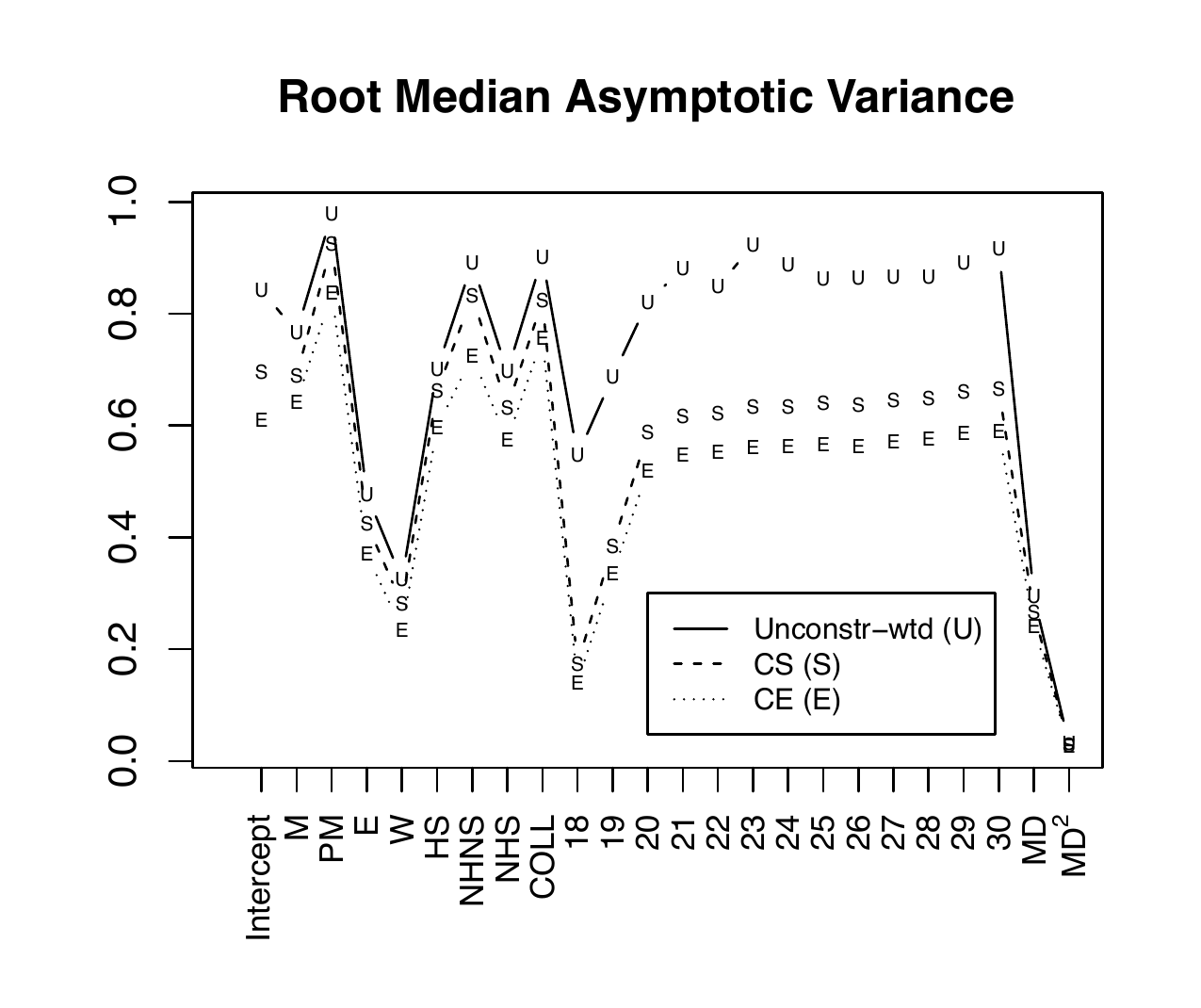}}
\caption{Parameter estimates and their asymptotic standard errors for PSID data.}
\label{fig:boot}
\end{center}
\end{figure}

The estimates $\h{\theta}_{PL}$, $\h{\theta}_{CS}$ and $\h{\theta}_{CE}$ of $\theta$ and their estimated asymptotic variances were calculated for each $10000$ de-clustered datasets.
The median of these parameter estimates and square root of the median of their estimated asymptotic variance over are presented in Figure \ref{fig:boot}.  
From the figure it is clear that on typically $\h{\theta}_{CE}$ is more efficient than both $\h{\theta}_{PL}$ and $\h{\theta}_{CS}$.
Note that while $\pi\ne\bp$, 
$\hat{\theta}_{CE}$
turns out to preform better than $\hat{\theta}_{CS}$.  Typically, the
standard errors for the age coefficients are about $10\%$ lower
for our CE estimator than the CS estimator.


\section{\bf Discussion}
In this article we present a new method to include design weights in an empirical likelihood based estimation procedure.  We also incorporate population level information in statistical modeling based on sample data.
Typically, in sample surveys, observations are selected with unequal probabilities due to purposive ``oversampling'', clustering, stratification, post-stratification, attrition and other non-response adjustments.
   For such surveys, the observed distribution of the sampled observations are different from their distribution in the population.  We adapt a parametric conditional likelihood (\citet{pfeffermann_1998}) to empirical likelihood and include population level information in the analysis.  
  Information about the model and the population are introduced as equality restrictions through estimating equations. 
  The parameter estimates are obtained by maximizing the empirical likelihood under these constraints by a two-step procedure.  The product of the weights can be interpreted as a non-parametric likelihood of the sample under the true population distribution.  
We assume that the sampling weights contain all information about the design. The expectation of these sampling weights conditional on the observed variables are used in the analysis.  

It is known that empirical likelihood and the estimators based on them have many desirable properties. \citet{owenbook} shows that the corresponding Wilk's statistic has an asymptotic Chi-squared limit for i.i.d. observations.  
Similar results follow for various kinds of dependence as well. 
 The constrained empirical likelihood can be expressed as a profile likelihood for $\theta$. \citet{Qinlawless1} show that, under standard regularity conditions, $\h{\theta}$ is asymptotically unbiased and normally distributed. 
\citet{mshren2} and \citet{chaudhuri_handcock_rendall_2008} show that it is beneficial to include available population level information in statistical modeling.  Such information is guaranteed to reduce the standard error of the estimates. 
As we show in Section 5, the two-step estimator used by
\citet{chaudhuri_handcock_rendall_2008} can be adapted to obtain
estimates.  Analytic expressions of the asymptotic standard
errors of the estimates are also known.

For empirical likelihood based estimators, one does not need to specify a parametric form for the likelihood. They are therefore more flexible and avoid unnecessary assumptions on the distribution of the design variables.  
For a fully parametric approach one needs to specify a parametric
candidate for $F^0$.  This is usually difficult and in practice
the model can be misspecified.  
The proposed estimator requires specification of two models.  One
for the response of interest and the auxiliary variables ($\psi_{\theta}(Y,A)$),
the other for the sampling probabilities 
($E_{\mc{P}}\left[\pi\mid V\right]$).  
This can often be done based on substantive knowledge.
Even if a parametric model for $F^0$ is correctly specified,
$\h{\theta}_{CE}$ is almost as efficient as the corresponding
CMLE.
In fact, if the underlying distribution is misspecified, our CE can be more efficient than the corresponding CMLE. For illustrative examples, we refer to \citet*{chaudhuri_drton_richardson_biom_2007}.

Empirical likelihood based methodologies have huge computational and implementational advantages over the corresponding constrained maximum likelihood estimators.  Direct non-linear equality constraints on the parameters often make computation infeasible. 
Empirical likelihood based methods put linear constraints on the weights, which can be implemented quite easily.
Furthermore, the parametric conditional likelihood involves a
difficult high-dimensional integral which our method avoids by
replacing $\Bp$ by $\sum\bp_iw_i$ in (\ref{eq:logemplik}).

The estimation of the weights in the CE requires a
constrained maximization problem to be solved. This can be achieved using 
the algorithms in \citet{owenbook} and \citet{chen_sitter_wu_2002}.
For generalized linear models, both have been implemented in
open-source software developed by the authors (BLINDED) 
We will make the methods developed in this paper available in this package.


Our estimator differs from the 
minimum divergence estimator (\citet{chen_sitter_1999}).
Our estimator incorporates the weights in the constraints while CS incorporates
them in their likelihood (c.f. \eqref{eq:esthstar}).
For large samples, our estimator is close to the correct likelihood.
For biased sampling designs, the CS likelihoods will tend to be further from the
super-population likelihoods than our CE likelihoods.
Our estimator is likely to be more efficient than the CS.
Also the design information is incorporated through
the conditional expectation of the sampling probabilities.  
This can be estimated even when the inclusion probabilities are not known.
This makes our estimator different from the estimator used by \citet{kim_2009}.

Our estimator is related to the inverse probability weighted
general Horvitz-Thompson type estimator.  In fact, without any
population level restriction, the parameter estimates are
obtained by solving the score equations weighted by inverse of
the conditional expectation of weights.  This justifies the
Horvitz-Thompson estimator with random weights and shows that it
can be derived from a likelihood perspective.

The relation between our estimator and the inverse probability
weighted general Horvitz-Thompson estimator is particularly
interesting with respect to model misspecification. It is known
that Horvitz-Thompson estimators are usually robust against model
misspecification.  It is an intriguing possibility that our
estimate inherits a part of this robustness as well.  The
robustness against model misspecification is particularly
beneficial, since it is often difficult to specify correct
distributions for the design variables.



The conditional empirical likelihood can be used as a likelihood
in Bayesian procedures.  In particular, this may be applied to
Bayesian analysis in problems in sample surveys, small area
estimation, epidemiology, case-control studies, among others.
     
\appendix
\section{Proofs}\label{sec:proof}
In this section we present the proofs of the theorems.\\  
\noindent{\bf Proof of Lemma \ref{lem:a1}}\begin{proof}
Recall that $E_{\mc{P}}\left[I_S\mid D_{\mc{P}}\right]=\pi_S$.  Now Assumption $1$ implies
\[\pi\left(S,D_{\mc{P}}\right)=E_{\mc{P}}\left[\pi_S\mid D_{\mc{P}}\right]=E_{\mc{P}}\left[E_{\mc{P}}\left[I_S\mid D_{\mc{P}}\right]\mid D_{\mc{P}}\right]=E_{\mc{P}}\left[I_S\mid D_{\mc{P}}\right]=\pi_S.\] The other side is immediate.
\end{proof}
\noindent{\bf Proof of Lemma \ref{lem:a2}}\begin{proof}
$1.$ Using $\pi_S=\pi\left(S,D_{\mc{P}}\right)$,  $E_{\mc{P}}\left[I_S\mid \pi_S,D_{\mc{P}}\right]=E_{\mc{P}}\left[I_S\mid D_{\mc{P}}\right]=\pi_S$.  This means 
\begin{equation}\label{eq:pis}
E_{\mc{P}}\left[I_S\mid \pi_S\right]=E_{\mc{P}}\left[E_{\mc{P}}\left[I_S\mid \pi_S,D_{\mc{P}}\right]\mid \pi_S\right]=E_{\mc{P}}\left[E_{\mc{P}}\left[I_S\mid D_{\mc{P}}\right]\mid \pi_S\right]=\pi_S.\end{equation}

\noindent $2.$ Clearly  $Pr_{\mc{P}}\left[I_S=1\mid \pi_S,D_{\mc{P}}\right]=\pi_S$.  Since $I_S$ is binary, its conditional distribution given $\pi_S$ and $D_{\mc{P}}$ is a function of $\pi_S$ only.  So from the definition of conditional independence \citep{lau} the result follows.
\end{proof}


\noindent{\bf Proof of Lemma \ref{lem:conda2}}\begin{proof} From \citet[page 29]{lau} it can be shown that, $\cind{I_S}{\left(X_{\mc{P}}, X_{\mc{P}}, D_{\mc{P}}\right)}{\pi_S}$ is equivalent to $\cind{I_S}{D_{\mc{P}}}{\pi_S}$ and 
$\cind{I_S}{\left(X_{\mc{P}}, X_{\mc{P}}\right)}{\left(\pi_S,D_{\mc{P}}\right)}$.  From Lemma \ref{lem:a1}, under Assumption $1$, the second conditional independence relationship is equivalent to $\cind{I_S}{\left(X_{\mc{P}}, X_{\mc{P}}\right)}{D_{\mc{P}}}$.   
\end{proof}

\vskip -7pt
\noindent{\bf Proof of Lemma \ref{lem:conda3}}\begin{proof}  The proofs follow from \citet[page 29]{lau}. We only present a sketches.\\
\noindent{$1.$} Follows from Assumption $2$.  \\
\noindent{$2.$} From Assumption $2$, it follows that $\cind{I_S}{\left(X_{\mc{P}}, X_{\mc{P}}\right)}{\left(\pi_S,D_{\mc{P}}\right)}$ holds.  This together with Assumption $1$ completes the proof.\\
\noindent{$3.$} This statement follows from $2.$ above.
\end{proof}

\noindent{\bf Proof of Theorem \ref{thm:FCE}}
\begin{proof}
Consider the objective function in \eqref{eq:logemplik}. Clearly for an extremum $0=\partial L/\partial w_i=(1/w_i)-\{n\bp_i/(\sum^n_{i=1}w_i\bp_i)\}$. This implies $w_i=(\sum^n_{i=1}w_i\bp_i)/(n\bp_i)$.
Now from $\sum^n_{i=1}w_i=1$ we obtain $\sum^n_{i=1}w_i\bp_i/n=1/\sum^n_{i=1}1/\bp_i$.  Thus $w_i=(1/\bp_i)/\sum^n_{i=1}1/\bp_i$ and the result follows.
\end{proof} 
\vskip -12pt
\noindent{\bf Proof of Lemma \ref{lem:reint}}\begin{proof}
Following \citet{owenbook} it can be shown that $\h{w}^{\star}_i=\left\{n(1+\xi^{\star}h_i/\bp_i)\right\}^{-1}$ with $(1+\xi^{\star}h_i/\bp_i)>n^{-1}$ for all $i=1,2,\ldots,n$, where $\xi^{\star}$ is the unique optimal value of the Lagrange multiplier $\xi$.

Further $\xi^{\star}$ satisfies,
$\sum^n_{i=1}(h_i/\bp_i)/(1+\xi^{\star} h_i/\bp_i)=0$.  This by uniqueness implies $\kappa=\xi^{\star}$.

Now by denoting $\h{\Bp}=\sum^n_{i=1}\bp_i\h{w}_{CEi}$ and from \eqref{eq:whatcs} by comparing $\h{w}^{\star}_i$ and $\h{w}_{CEi}$ we notice that 
\begin{equation}\label{eq:hwstar}
\h{w}^{\star}_i=\frac{\bp_i\h{w}_{CEi}}{\h{\Bp}}.
\end{equation}

Now summing over $i$ and noting that $\sum^n_{i=1}\h{w}_{CEi}=1$ it follows that $\h{\Bp}=\left(\sum^n_{i=1}\h{w}^{\star}_i/\bp_i\right)^{-1}$.  Substituting these results in \eqref{eq:whatcs} it follows that
$\h{w}_{CEi}=(\h{w}^{\star}_i/\bp_i)/\sum^n_{i=1}(\h{w}^{\star}_i/\bp_i)$.
Further note that by the non negativity of $w^{\star}_i$  
\begin{equation}
\h{\Bp}^{-1}=\sum^n_{i=1}\frac{\h{w}^{\star}_i}{\bp_i}=\frac{1}{n}\sum^n_{i=1}\frac{1}{\bp_i+\kappa h_i}\ge\frac{1}{n\left(\bp_i+\kappa h_i\right)}.
\end{equation}
for all $i=1,2,\ldots,n$, which is the restriction on $\h{w}_{CEi}$ in \eqref{eq:whatcs}. 
\end{proof}
\noindent{\bf Proof of Corollary \ref{cor:CS}}
\begin{proof}The condition means, $G^{-1}\left(K_1H^{-1}_1H_2-K_2\right)H^{-1}_1=0$.  This implies $K_1H_1^{-1}-K_2H^{-1}_2=0$. Now by substituting, $K_2H^{-1}_2$ for $K_1H_1^{-1}$ in the expression of $\mc{V}_{CS}$, the result follows. 
\end{proof} 
\noindent{\bf Proof of Theorem \ref{thm:stdgain}}
\begin{proof}
Since $\bp=\pi$, $\mG=G$, $\mG^{\star}=G^{\star}$, $\mK_2=K_2$ and $\mH_2=H_2$, simple algebraic manipulation (see the supplement) shows that:
\begin{equation}
G(\mathcal{V}_{CS}-\mathcal{V}_{CE})G^{T}=\left(K_2H_2^{-1}-K_1H^{-1}_1\right)H_2\left(H^{-1}_2K^T_2-H^{-1}_1K^T_1\right).
\end{equation} 
Furthermore, $\h{\theta}^{(N)}_{CS}$ and $\h{\lambda}^{(N)}$ are asymptotically uncorrelated implies $K_2H_2^{-1}-K_1H^{-1}_1=0$.  From this the results clearly follow.
\end{proof}

\section{Asymptotic properties of the estimators}
In this section we discuss the asymptotic properties of the two parameter estimates of $\theta$ obtained from the two empirical likelihood based methods (CE and CS) under the true distribution $F^0$ in the population.  
We only consider two-step estimation and show that as $N\rightarrow\infty$, both  $\h{\theta}_{CS}$ and $\h{\theta}_{CE}$ are strongly consistent and asymptotically normal. They have different asymptotic covariance matrices, which we express analytically.  

We first discuss notation and specify the assumptions.  Let us denote :
\begin{align}
f_1\left(v,d,\theta,\lambda\right)&=\frac{d}{1+\lambda h\left(v,\gamma\right)}\left(\psi_{\theta}(y,a),h\left(v,\gamma\right)\right)\\
f_2\left(v,\bp,\theta,\kappa\right)&=\frac{1}{\bp+\kappa h\left(v,\gamma\right)}\left(\psi_{\theta}(y,a),h\left(v,\gamma\right)\right)
\end{align}

Suppose $\theta_0$ is the true value of $\theta$.  Following \citet{Qinlawless1} and \citet{serfling1} we make the following assumptions.
\begin{list}{}{}
\item[A.$1.$] We assume that both $f_1\left(v_i,d_i,\theta,\lambda\right)$, $1\le i\le N$ and $f_2\left(v_i,\bp_i,\theta,\kappa\right)$, $1\le i\le N$ are i.i.d. random vectors for any $\theta$ and $\lambda$.  
\item[A.$2.$] Suppose for all $d$ and $\bp$, $E_{\mc{P}}\left[\fd\right]=E_{\mc{P}}\left[\fp\right]=0$.
\item[A.$3.$] Both Jacobians $\dfd$ and $\dfp$ and Hessians \\$\partial^2 f_1\left(v,d,\theta,\lambda\right)/\partial^2\left(\theta,\lambda\right)$ and $\partial^2 f_2\left(v,\bp,\theta,\kappa\right)/\partial^2\left(\theta,\kappa\right)$ exists for all $\theta$, $\lambda$ and $\kappa$ and the Jacobian matrices are continuous in the neighbourhood of the true value $\left(\theta_0,0\right)$.

\item[A.$4(a)$] With $\norm{~\cdot~}$ denoting the Euclidean norm, suppose that $\norm{\dfd}$, $\norm{f_1\left(v,d,\theta,\lambda\right)}^3$ and $\norm{\partial^2 f_1\left(v,d,\theta,\lambda\right)/\partial^2\left(\theta,\lambda\right)}$ are bounded by $\mathfrak{G}(v,d)$ 
for some integrable function $\mathfrak{G}$ in the neighbourhood of $\left(\theta_0,0\right)$.
 
\item[A.$4(b)$] $\norm{\dfp}$, $\norm{f_2\left(v,\bp,\theta,\kappa\right)}^3$ and $\norm{\partial^2 f_2\left(v,\bp,\theta,\kappa\right)/\partial^2\left(\theta,\kappa\right)}$ are bounded by $\mathfrak{H}(v,\bp)$ for some integrable function $\mathfrak{H}$ in the neighbourhood of $\left(\theta_0,0\right)$.

\item[A.$5.$] Both $E_{\mc{P}}\left[\fd\fd^T\right]$ and  $E_{\mc{P}}\left[\fp\fp^T\right]$ are positive definite matrices.
 
\item[A.$6.$] Both $E_{\mc{P}}\left[\dfdt\right]$ and $E_{\mc{P}}\left[\dfpt\right]$ have full ranks.  
\end{list}

The next two proofs closely follow \citet{chaudhuri_handcock_rendall_2008}.  We only present the sketch of the arguments here.  The details can be found in the above reference.

\begin{thm}\label{thm:CS2}
Under our assumptions $A.1-A.4(a)$ and $A.5-A.6$, almost surely the equation $\sum^N_{i=1}f_1\left(v_i,d_i,\theta,\lambda\right)=0$ admits a sequence of solutions $(\h{\theta}^{(N)}_{CS},\h{\lambda}^{(N)})$ such that  

\begin{enumerate}{}{}
\item $(\h{\theta}^{(N)}_{CS},\h{\lambda}^{(N)})$ $\longrightarrow$ $(\theta_0,0)$ as $N\rightarrow\infty$,
\item $N^{1/2}(\h{\theta}^{(N)}_{CS}-\theta_0)\Rightarrow N\left(0,\mathcal{V}_{CS}\right)$ distribution, where\\
$\mathcal{V}_{CS}=G^{-1}\left(G^{\star}-K_1H^{-1}_1K^T_2-K_2H^{-1}_1K^T_1+K_1H^{-1}_1H_2H^{-1}_1K^T_1\right)\left(G^T\right)^{-1}$,
\item $N^{1/2}\h{\lambda}^{(N)}\Rightarrow N\left(0,H^{-1}_1H_2H^{-1}_1\right)$ distribution,
\item Asymptotic covariance of $\h{\theta}^{(N)}_{CS}$ and $\h{\lambda}^{(N)}$ is given by $G^{-1}\left(K_1H^{-1}_1H_2-K_2\right)H^{-1}_1$.
\end{enumerate}
\begin{proof}
Note that:
\begin{equation}
(-1)\cdot\frac{\partial f_1\left(v_i,d_i,\theta,\lambda\right)}{\partial\left(\theta,\lambda\right)}=\begin{pmatrix}
-\frac{d_i\psi^{\prime}\left(v_i,\theta\right)}{1+\lambda h\left(v_i,\gamma\right)}&\frac{d_i\psi\left(v_i,\theta\right)h\left(v_i,\gamma\right)}{\left\{1+\lambda h\left(v_i,\gamma\right)\right\}^2}\\
0&\frac{d_ih^2\left(v_i,\gamma\right)}{\left\{1+\lambda h\left(v_i,\gamma\right)\right\}^2}
\end{pmatrix}.
\end{equation}
Thus 
\begin{equation}
(-1)\cdot E_{\mc{P}}\left\{\left.\frac{\partial f_1\left(v_1,d_1,\theta,\lambda\right)}{\partial\left(\theta,\lambda\right)}\right|_{\theta=\theta_0,\lambda=0}\right\}=\begin{pmatrix}
-G&K_1\\
0&H_1
\end{pmatrix}.
\end{equation}
Further 
\begin{equation}
\hbox{Var}_{\mc{P}}\left\{f_1\left(v_1,d_1,\theta_0,0\right)\right\}=\begin{pmatrix}
G^{\star}&K_2\\
K^T_2&H_2
\end{pmatrix}.
\end{equation}
Now by expanding $\sum^N_{i=1}f_1\left(v_1,d_1,\h{\theta}^{(N)}_{CS},\h{\lambda}^{(N)}\right)/N$ around $\left(\theta_0,0\right)$, under the assumptions, the results can be shown via standard techniques.

In particular, $\sqrt{N}\left((\h{\theta}^{(N)}_{CS}-\theta_0),\h{\lambda}^{(N)}\right)^T$ converges to a normal distribution with covariance matrix:
%
\begin{align}
&\begin{pmatrix}
-G&K_1\\
0&H_1
\end{pmatrix}^{-1}\begin{pmatrix}
G^{\star}&K_2\\
K^T_2&H_2
\end{pmatrix}\begin{pmatrix}
-G^T&0\\
K^T_1&H_1
\end{pmatrix}^{-1}\nonumber\\
=&{\tiny\begin{bmatrix}
G^{-1}\left\{G^{\star}-K_1H^{-1}_1K^T_2-K_2H^{-1}_1K^T_1+K_1H^{-1}_1H_2H^{-1}_1K^T_1\right\}\left(G^T\right)^{-1}&G^{-1}\left(K_1H^{-1}_1H_2-K_2\right)H^{-1}_1\\
H^{-1}_1\left(K^T_2-H^{-1}_2H_1K^T_1\right)G^{-1}_1&H^{-1}_1H_2H^{-1}_1
\end{bmatrix}\nonumber}
\end{align}
\end{proof}
\end{thm}
\begin{thm}\label{thm:CL}
Under our assumptions $A.1-A.3$ and $A.4(b)-A.6$, almost surely the equation $\sum^N_{i=1}f_2\left(v_i,\bp_i,\theta,\kappa\right)=0$ admits a sequence of solutions $(\h{\theta}^{(N)}_{CE},\h{\kappa}^{(N)})$ such that  

\begin{enumerate}{}{}
\item $(\h{\theta}^{(N)}_{CE},\h{\kappa}^{(N)})$ $\longrightarrow$ $(\theta_0,0)$ as $N\rightarrow\infty$,
\item $N^{1/2}(\h{\theta}^{(N)}_{CE}-\theta_0)\Rightarrow N\left(0,\mathcal{V}_{CE}\right)$ distribution, where $\mathcal{V}_{CE}=\mG^{-1}\left(\mG^{\star}-\mK_2\mH^{-1}_2\mK^T_2\right)\left(\mG^T\right)^{-1}$,
\item $N^{1/2}\h{\kappa}^{(N)}\Rightarrow N\left(0,\mH^{-1}_2\right)$ distribution,
\item $\h{\theta}^{(N)}_{CE}$ and $\h{\kappa}^{(N)}$ are asymptotically independent.
\end{enumerate}
\begin{proof}
The proof is similar to that of Theorem \ref{thm:CS}.  Note that :
\begin{equation}
(-1)\cdot\frac{\partial f_2\left(v_i,\bp_i,\theta,\kappa\right)}{\partial\left(\theta,\kappa\right)}=\begin{pmatrix}
-\frac{\psi^{\prime}\left(v_i,\theta\right)}{\bp_i+\kappa h\left(v_i,\gamma\right)}&\frac{\psi\left(v_i,\theta\right)h\left(v_i,\gamma\right)}{\left\{\bp_i+\kappa h\left(v_i,\gamma\right)\right\}^2}\\
0&\frac{h^2\left(v_i,\gamma\right)}{\left\{\bp_i+\kappa h\left(v_i,\gamma\right)\right\}^2}
\end{pmatrix}.
\end{equation}
Thus
\begin{equation}
(-1)\cdot E_{\mc{P}}\left\{\left.\frac{\partial f_2\left(v_i,\bp_i,\theta,\kappa\right)}{\partial\left(\theta,\kappa\right)}\right|_{(\theta=\theta_0,\kappa=0)}\right\}=\begin{pmatrix}
-\mG&\mK_2\\
0&\mH_2
\end{pmatrix}.
\end{equation}
Similar to Theorem \ref{thm:CS} we know that:
\begin{equation}
\hbox{Var}_{\mc{P}}\left\{f_2\left(v_i,\bp_i,\theta_0,0\right)\right\}=\begin{pmatrix}
\mG^{\star}_2&\mK_2\\
\mK^T_2&\mH_2
\end{pmatrix}
\end{equation}
Now from the expansion of $\sum^N_{i=1}f_2\left(v_i,\bp_i,\h{\theta}^{(N)}_{CE},\h{\kappa}^{(N)}\right)/N$ in the neighbourhood of $\left(\theta_0,0\right)$ as before, the results follow.

Furthermore, as before the asymptotic variance of $\sqrt{N}\left((\h{\theta}^{(N)}_{CE}-\theta_0),\h{\kappa}^{(N)}\right)^T$ is given by:
\begin{equation*}
\begin{pmatrix}
-\mG&\mK_2\\
0&\mH_2
\end{pmatrix}^{-1}
\begin{pmatrix}
\mG^{\star}_2&\mK_2\\
\mK^T_2&\mH_2
\end{pmatrix}
\begin{pmatrix}
-\mG^T&0\\
\mK^T_2&\mH_2
\end{pmatrix}^{-1}=\begin{pmatrix}
\mG^{-1}\left\{\mG^{\star}_2-\mK^T_2\mH^{-1}_2\mK_2\right\}\left(\mG^T\right)^{-1}&0\\
0&\mH^{-1}_2
\end{pmatrix}
\end{equation*}
\end{proof}
\end{thm}
The following theorem shows that if $\bp=\pi$, $\h{\theta}_{CE}$ is more efficient than $\h{\theta}_{CS}$.  There is a sketch of the proof in the Appendix of the main article.  We present the details below.
\begin{thm}\label{thm:stdgain2}
If $\bp_i=\pi_i$, for all $i=1,2,\ldots,N$, the asymptotic standard error of $\h{\theta}_{CS}$ is larger than $\h{\theta}_{CE}$.
\begin{proof}
Since $\bp=\pi$, $\mG=G$, $\mG^{\star}=G^{\star}$, $\mK_2=K_2$ and $\mH_2=H_2$. 
\begin{align}
G(\mathcal{V}_{CS}-\mathcal{V}_{CE})G^{T}&=K_2H^{-1}_2K^T_2-K_1H^{-1}_1K^T_2-K_2H^{-1}_1K^T_1+K_1H^{-1}_1H_2H^{-1}_1K^T_1\nonumber\\
&=K_2\left(H^{-1}_2K^T_2-H^{-1}_1K^T_1\right)+K_1H^{-1}_1\left(H_2H^{-1}_1K^T_1-K^T_2\right)\nonumber\\
&=K_2\left(H^{-1}_2K^T_2-H^{-1}_1K^T_1\right)+K_1H^{-1}_1H_2\left(H^{-1}_1K^T_1-H^{-1}_2K^T_2\right)\nonumber\\
&=\left(K_2-K_1H^{-1}_1H_2\right)\left(H^{-1}_2K^T_2-H^{-1}_1K^T_1\right)\nonumber\\
&=\left(K_2H_2^{-1}-K_1H^{-1}_1\right)H_2\left(H^{-1}_2K^T_2-H^{-1}_1K^T_1\right)\nonumber
\end{align}
Clearly $G(\mathcal{V}_{CS}-\mathcal{V}_{CE})G^{T}$ is a non-negative definite matrix. So $\mathcal{V}_{CS}-\mathcal{V}_{CE}$ is non-negative definite as well.
\end{proof}
\end{thm}


\begin{thebibliography}{}

\bibitem[\protect\citeauthoryear{Beaumont}{Beaumont}{2008}]{beaumont_2008}
Beaumont, J.-F. (2008).
\newblock A new approach to weighting and inference in sample surveys.
\newblock {\em Biometrika\/}~{\em 95\/}(3), 539--553.

\bibitem[\protect\citeauthoryear{BLINDED}{BLINDED}{2012}]{blinded}
BLINDED (2012).
\newblock Technical report.

\bibitem[\protect\citeauthoryear{Breslow and Wellner}{Breslow and
  Wellner}{2006}]{breslow_wellner_2006}
Breslow, N. and J.~Wellner (2006).
\newblock Weighted likelihood for semiparametric models and two -phase
  stratified samples, with application to cox regression.
\newblock {\em Scand. J. Statist\/}~{\em 34}, 86--102.

\bibitem[\protect\citeauthoryear{Chambers}{Chambers}{2003}]{chambers_2003}
Chambers, R.~L. (2003).
\newblock Introduction to part a.
\newblock In {\em Analysis of Survey data}, pp.\  13--28.

\bibitem[\protect\citeauthoryear{Chambers, Dorfman, and Sverchkov}{Chambers
  et~al.}{2003}]{chambers_dorfman_sverchkov_2003}
Chambers, R.~L., A.~H. Dorfman, and M.~Y. Sverchkov (2003).
\newblock Nonparametric regression with complex survey data.
\newblock In {\em Analysis of survey data ({S}outhampton, 1999)}, Wiley Ser.
  Surv. Methodol., pp.\  151--174. Chichester: Wiley.

\bibitem[\protect\citeauthoryear{Chaudhuri, Drton, and Richardson}{Chaudhuri
  et~al.}{2007}]{chaudhuri_drton_richardson_biom_2007}
Chaudhuri, S., M.~Drton, and T.~S. Richardson (2007).
\newblock Estimation of a covariance matrix with zeros.
\newblock {\em Biometrika\/}~{\em 94\/}(1), 199--216.

\bibitem[\protect\citeauthoryear{Chaudhuri, Handcock, and Rendall}{Chaudhuri
  et~al.}{2008}]{chaudhuri_handcock_rendall_2008}
Chaudhuri, S., M.~S. Handcock, and M.~S. Rendall (2008).
\newblock Generalized linear models incorporating population level information:
  an empirical-likelihood-based approach.
\newblock {\em Journal of the Royal Statistical Society series B\/}~{\em 70},
  311--328.

\bibitem[\protect\citeauthoryear{Chen and Sitter}{Chen and
  Sitter}{1999}]{chen_sitter_1999}
Chen, J. and R.~R. Sitter (1999).
\newblock A pseudo empirical likelihood approach to the effective use of
  auxiliary information in complex surveys.
\newblock {\em Statist. Sinica\/}~{\em 9\/}(2), 385--406.

\bibitem[\protect\citeauthoryear{Chen, Sitter, and Wu}{Chen
  et~al.}{2002}]{chen_sitter_wu_2002}
Chen, J., R.~R. Sitter, and C.~Wu (2002).
\newblock Using empirical likelihood methods to obtain range restricted weights
  in regression estimators for surveys.
\newblock {\em Biometrika\/}~{\em 89\/}(1), 230--237.

\bibitem[\protect\citeauthoryear{Chen and Qin}{Chen and
  Qin}{1993}]{chen_qin_1993}
Chen, J.~H. and J.~Qin (1993).
\newblock Empirical likelihood estimation for finite populations and the
  effective usage of auxiliary information.
\newblock {\em Biometrika\/}~{\em 80\/}(1), 107--116.

\bibitem[\protect\citeauthoryear{Fu, Wang, and Wu}{Fu
  et~al.}{2008}]{fu_wang_wu_2008}
Fu, Y., X.~Wang, and C.~Wu (2008).
\newblock Weighted empirical likelihood inference for multiple samples.
\newblock {\em Journal of Statistical Planning and Inference\/}.

\bibitem[\protect\citeauthoryear{Fuller}{Fuller}{2009}]{fuller_2009}
Fuller, W.~A. (2009).
\newblock {\em Sampling Statistics}.
\newblock Wiley and Sons.

\bibitem[\protect\citeauthoryear{Gill, Vardi, and Wellner}{Gill
  et~al.}{1988}]{gill_vardi_wellner_1988}
Gill, R.~D., Y.~Vardi, and J.~A. Wellner (1988).
\newblock Large sample theory of empirical distributions in biased sampling
  models.
\newblock {\em Ann. Statist.\/}~{\em 16\/}(3), 1069--1112.

\bibitem[\protect\citeauthoryear{Glenn and Zhao}{Glenn and
  Zhao}{2007}]{glenn_zhao_2007}
Glenn, N. and Y.~Zhao (2007).
\newblock Weighted empirical likelihood estimates and their robustness
  properties.
\newblock {\em Computational Statistics \& Data Analysis\/}~{\em 51},
  5130--5141.

\bibitem[\protect\citeauthoryear{Godambe}{Godambe}{1975}]{godambe75}
Godambe, V.~P. (1975).
\newblock A reply to my critics.
\newblock {\em Sankhya, Series C\/}~{\em 37}, 53--76.

\bibitem[\protect\citeauthoryear{Handcock, Huovilainen, and Rendall}{Handcock
  et~al.}{2000}]{mshren1}
Handcock, M.~S., S.~M. Huovilainen, and M.~S. Rendall (2000).
\newblock Combining registration-system and survey data to estimate birth
  probabilities.
\newblock {\em Demography\/}~{\em 37\/}(2), 187--192.

\bibitem[\protect\citeauthoryear{Handcock, Rendall, and Cheadle}{Handcock
  et~al.}{2005}]{mshren2}
Handcock, M.~S., M.~S. Rendall, and J.~E. Cheadle (2005).
\newblock Improved regression estimation of a multivariate relationship with
  population data on the bivariate relationship.
\newblock {\em Sociological Methodology\/}~{\em 35\/}(1), 291--334.

\bibitem[\protect\citeauthoryear{Hartley and Rao}{Hartley and
  Rao}{1968}]{hartley_rao_1968}
Hartley, H.~O. and J.~N.~K. Rao (1968).
\newblock A new estimation theory for sample surveys.
\newblock {\em Biometrika\/}~{\em 55\/}(3), 547--557.

\bibitem[\protect\citeauthoryear{Hartley and Rao}{Hartley and
  Rao}{1969}]{hartley_rao_1969}
Hartley, H.~O. and J.~N.~K. Rao (1969).
\newblock A new estimation theory for sample surveys, ii.
\newblock In {\em New Developments in Survay Sampling: A Symposium on the
  Foundations of Survey Sampling held at the University of North Carolina,
  Chapel Hill, North Carolina}, pp.\  147--169. New York: Wiley.

\bibitem[\protect\citeauthoryear{Hartley}{Hartley}{1975}]{hartleyrao75}
Hartley, H. O.~Rao, J. N.~K. (1975).
\newblock Some comments on labels: A rejoinder to the section of godambe's
  paper, 'a reply to my critics'.
\newblock {\em Sankhya, Series C\/}~{\em 37}, 163--170.

\bibitem[\protect\citeauthoryear{Imbens and Lancaster}{Imbens and
  Lancaster}{1994}]{imbens2}
Imbens, G.~W. and T.~Lancaster (1994).
\newblock Combining micro and macro data in microeconomic models.
\newblock {\em Review of Economic Studies\/}~{\em 61}, 655--380.

\bibitem[\protect\citeauthoryear{{Institute for Social Research}}{{Institute
  for Social Research}}{2010}]{PSID2010}
{Institute for Social Research} (2010).
\newblock Panel study of income dynamics, public use dataset [machine-readable
  data file and documentation].
\newblock Technical report, Survey Research Center, University of Michigan, Ann
  Arbor, MI.
\newblock Produced and distributed by the Institute for Social Research, Survey
  Research Center, University of Michigan, Ann Arbor, MI.

\bibitem[\protect\citeauthoryear{Kim}{Kim}{2009}]{kim_2009}
Kim, J.~K. (2009).
\newblock Calibration estimation using empirical likelihood in survey sampling.
\newblock {\em Statist. Sinica\/}~{\em 19\/}(1), 145--157.

\bibitem[\protect\citeauthoryear{Krieger and Pfeffermann}{Krieger and
  Pfeffermann}{1992}]{krieger_pfeffermann_1992}
Krieger, A.~M. and D.~Pfeffermann (1992).
\newblock Maximum likelihood estimation from complex sample surveys.
\newblock {\em Survey Methodology\/}~{\em 18\/}(2), 225--239.

\bibitem[\protect\citeauthoryear{Lauritzen}{Lauritzen}{1996}]{lau}
Lauritzen, S.~L. (1996).
\newblock {\em Graphical Models}.
\newblock Oxford: Clarendon Press.

\bibitem[\protect\citeauthoryear{McCullagh and Nelder}{McCullagh and
  Nelder}{1989}]{mccneldrbook}
McCullagh, P. and J.~Nelder (1989).
\newblock {\em Generalised Linear Models}.
\newblock Chapman\& Hall/CRC.

\bibitem[\protect\citeauthoryear{Owen}{Owen}{2001}]{owenbook}
Owen, A. (2001).
\newblock {\em Empirical Likelihood}.
\newblock Chapman\& Hall/CRC.

\bibitem[\protect\citeauthoryear{Patil and Rao}{Patil and
  Rao}{1978}]{rao_patil_1978}
Patil, G.~P. and C.~R. Rao (1978).
\newblock Weighted distributions and size-biased sampling with applications to
  wildlife populations and human families.
\newblock {\em Biometrics\/}~{\em 34\/}(2), 179--189.

\bibitem[\protect\citeauthoryear{Pfeffermann, Krieger, and Rinott}{Pfeffermann
  et~al.}{1998}]{pfeffermann_1998}
Pfeffermann, D., A.~M. Krieger, and Y.~Rinott (1998).
\newblock Parametric distributions of complex survey data under informative
  probability sampling.
\newblock {\em Statist. Sinica\/}~{\em 8\/}(4), 1087--1114.

\bibitem[\protect\citeauthoryear{Pfeffermann and Sverchkov}{Pfeffermann and
  Sverchkov}{1999}]{pfeffermann_sverchkov_1999}
Pfeffermann, D. and M.~Sverchkov (1999).
\newblock Parametric and semi-parametric estimation of regression models fitted
  to survey data.
\newblock {\em Sankhy\=a Ser. B\/}~{\em 61\/}(1), 166--186.

\bibitem[\protect\citeauthoryear{Pfeffermann and Sverchkov}{Pfeffermann and
  Sverchkov}{2003}]{pfeffermann_sverchkov_2003}
Pfeffermann, D. and M.~Sverchkov (2003).
\newblock Fitting generalized linear models under informative sampling.
\newblock In {\em Analysis of Survey data}, pp.\  175 -- 195. Chichester:
  Wiley.

\bibitem[\protect\citeauthoryear{Qin}{Qin}{1993}]{qin_1993}
Qin, J. (1993).
\newblock Empirical likelihood in biased sample problems.
\newblock {\em The Annals of Statistics\/}~{\em 21\/}(3), 1182--1196.

\bibitem[\protect\citeauthoryear{Qin and Lawless}{Qin and
  Lawless}{1994}]{Qinlawless1}
Qin, J. and J.~Lawless (1994).
\newblock Empirical likelihood and general estimating equations.
\newblock {\em The Annals of Statistics\/}~{\em 22}, 300--325.

\bibitem[\protect\citeauthoryear{Qin, Leung, and Shao}{Qin
  et~al.}{2002}]{qin_leung_shao_2002}
Qin, J., D.~Leung, and J.~Shao (2002).
\newblock Estimation with survey data under nonignorable nonresponse or
  informative sampling.
\newblock {\em J. Amer. Statist. Assoc.\/}~{\em 97\/}(457), 193--200.

\bibitem[\protect\citeauthoryear{Qin and Zhang}{Qin and
  Zhang}{2007}]{qin_zhang_2007}
Qin, J. and B.~Zhang (2007).
\newblock Empirical-likelihood-based inference in missing response problems and
  its application in observational studies.
\newblock {\em J. R. Stat. Soc. Ser. B Stat. Methodol.\/}~{\em 69\/}(1),
  101--122.

\bibitem[\protect\citeauthoryear{Rao and Wu}{Rao and Wu}{2008}]{rao_wu_2008}
Rao, J. N.~K. and C.~Wu (2008).
\newblock Empirical likelihood methods.
\newblock In R.~C. Pfeffermann~D. (Ed.), {\em Handbook of statistics, Sample
  Surveys: Inference and Analysis}, Volume 29B, pp.\  189--207. Elsevier.

\bibitem[\protect\citeauthoryear{Rendall, Admiraal, DeRose, DiGiulio, Handcock,
  and Racioppi}{Rendall et~al.}{2008}]{rendall_et_al_2008}
Rendall, M.~S., R.~Admiraal, A.~DeRose, P.~DiGiulio, M.~S. Handcock, and
  F.~Racioppi (2008).
\newblock Population constraints on pooled surveys in demographic hazard
  modelling.
\newblock {\em Statistical Methods and Applications\/}~{\em 17\/}(4), 519--539.

\bibitem[\protect\citeauthoryear{Schoen}{Schoen}{2005}]{schoen_2005}
Schoen, R. (2005).
\newblock Insights from parity status life tables for the 20th century u.s.
\newblock {\em Social Science Research\/}~{\em 35\/}(1), 29--39.

\bibitem[\protect\citeauthoryear{Scott}{Scott}{1977}]{scott_1977}
Scott, A.~J. (1977).
\newblock Some comments on the problem of randomisation in surveys.
\newblock {\em Sankhy$\bar{a}$ C\/}~{\em 39}, 1--9.

\bibitem[\protect\citeauthoryear{Serfling}{Serfling}{1980}]{serfling1}
Serfling, R.~J. (1980).
\newblock {\em Approximation Theorems of Mathematical Statistics}.
\newblock John Willey \& Sons.

\bibitem[\protect\citeauthoryear{Sugden and Smith}{Sugden and
  Smith}{1984}]{sugden_smith_1984}
Sugden, R.~A. and T.~M.~F. Smith (1984).
\newblock Ignorable and informative designs in survey sampling inference.
\newblock {\em Biometrika\/}~{\em 71\/}(3), 495--506.

\bibitem[\protect\citeauthoryear{Sullivan}{Sullivan}{2005}]{sullivan_2005}
Sullivan, R. (2005).
\newblock The age pattern of first-birth rates among u.s. women: The bimodal
  1990s.
\newblock {\em Demography\/}~{\em 42\/}(2), 259--273.

\bibitem[\protect\citeauthoryear{Tighe, Livert, and Saxe}{Tighe
  et~al.}{2010}]{tighe_livert_saxe_2010}
Tighe, E., D.~Livert, and L.~Saxe (2010).
\newblock Cross-survey analysis to estimate low-incidence religious groups.
\newblock {\em Socilogical Methods and Research\/}~{\em 39\/}(1), 56--82.

\bibitem[\protect\citeauthoryear{Vardi}{Vardi}{1985}]{vardi_1985}
Vardi, Y. (1985).
\newblock Empirical distributions in selection bias models.
\newblock {\em Ann. Statist.\/}~{\em 13\/}(1), 178--205.
\newblock With discussion by C. L. Mallows.

\bibitem[\protect\citeauthoryear{Wu}{Wu}{2004}]{wu_2004}
Wu, C. (2004).
\newblock Weighted empirical likelihood inference.
\newblock {\em Statist. Probab. Lett.\/}~{\em 66\/}(1), 67--79.

\bibitem[\protect\citeauthoryear{Wu and Rao}{Wu and Rao}{2006}]{wu_rao_2006}
Wu, C. and J.~N.~K. Rao (2006).
\newblock Pseudo-empirical likelihood ratio confidence intervals for complex
  surveys.
\newblock {\em Canad. J. Statist.\/}~{\em 34\/}(3), 359--375.

\end{thebibliography}

\end{document}